\documentclass[prb,superscriptaddress,showpacs,amsmath,amssymb,twocolumn]{revtex4-1}

\usepackage{xcolor}
\usepackage{graphicx}
\usepackage{bm}
\usepackage{hyperref}
\usepackage{units}
\usepackage[normalem]{ulem}

\newcommand{\dd}[1]{\mathrm{d}#1\,}
\newcommand{\DD}[1]{\mathrm{D}[#1]\,}
\newcommand{\avg}[1]{\langle{#1}\rangle}
\renewcommand{\Re}{\mathop{\mathrm{Re}}}

\DeclareMathOperator{\tr}{tr}
\newcommand{\sgn}{\mathop{\mathrm{sgn}}}
\renewcommand{\vec}[1]{\bm{#1}}

\definecolor{TTH-color}{named}{green}
\definecolor{RO-color}{named}{magenta}
\definecolor{PV-color}{rgb}{0.97,0.57,0.11}
\definecolor{JM-color}{RGB}{128,0,128}

\definecolor{TTH-color2}{named}{red}
\definecolor{RO-color2}{named}{blue}
\definecolor{PV-color2}{rgb}{0.87,0.47,0.01}
\definecolor{JM-color2}{RGB}{128,0,128}

\begin{document}

\title{Non-linear spin torque, pumping and cooling in superconductor/ferromagnet systems}

\author{Risto Ojaj\"arvi}
\email{risto.m.m.ojajarvi@jyu.fi}

\affiliation{University of Jyvaskyla, Department of Physics and Nanoscience Center, P.O. Box 35 (YFL), FI-40014 University of Jyv\"askyl\"a, Finland }
\author{Juuso Manninen}
\affiliation{Aalto University, Department of Applied Physics, Low Temperature Laboratory, P.O. Box 15100, FI-00076 AALTO, Finland}
\author{Tero T. Heikkil\"a}
\email{tero.t.heikkila@jyu.fi}
\affiliation{University of Jyvaskyla, Department of Physics and Nanoscience Center, P.O. Box 35 (YFL), FI-40014 University of Jyv\"askyl\"a, Finland }
\author{Pauli Virtanen}
\email{pauli.t.virtanen@jyu.fi}

\affiliation{University of Jyvaskyla, Department of Physics and Nanoscience Center, P.O. Box 35 (YFL), FI-40014 University of Jyv\"askyl\"a, Finland }

\date{\today}

\begin{abstract}
  We study the effects of the coupling between magnetization dynamics
  and the electronic degrees of freedom in a heterostructure
  of a metallic nanomagnet with dynamic magnetization coupled
  with a superconductor containing a steady spin-splitting field. We predict how this system exhibits a non-linear spin torque, which can be driven either with a temperature
  difference or a voltage across the interface. We
  generalize this notion to arbitrary magnetization precession by
  deriving a Keldysh action for the interface, describing
  the coupled charge, heat and spin transport in the presence of a
  precessing magnetization. We characterize
  the effect of superconductivity on the precession
  damping and the anti-damping torques. We also predict the full
  non-linear characteristic of the Onsager counterparts of the torque,
  showing up via pumped charge and heat currents. For the latter, we
  predict a spin-pumping cooling effect, where the magnetization dynamics can
  cool either the nanomagnet or the superconductor.
\end{abstract}


\maketitle

\section{Introduction}

The intriguing possibility to control magnetization dynamics by spin torque suggested
over two decades ago \cite{slonczewski1996-cde} and its
reciprocal counterpart \cite{johnson1987-tai,bauer2012-sc} of spin pumping
\cite{tserkovnyak2002-spm} have been widely studied in magnetic
systems. In such systems charge and spin transport are
closely linked and need to be treated on the same footing.
Recently there has also been increased interest in coupling
superconductors to magnets and finding out how superconductivity
affects the magnetization dynamics \cite{bell2008-sds,houzet2008-fjj,jeon2018-esp,yao2018-psd,jeon2019-ems,rogdakis2019-stp,morten2008-pea,skadsem2011-frv,adachi2017-spi,teber2010-tmd,richard2012-aci,holmqvist2014-sps,hammar2017-tsd,kato2019-mts,dutta2017thermoelectric}. 
On the other hand, recent work has shown that a combination of
magnetic and superconducting systems results in giant thermoelectric
effects
\cite{machon2013-nte,ozaeta2014-hte,silaev2015-lrs,bergeret2018-cne,heikkila2019-tes}
which couple charge and heat currents. These works
\cite{ozaeta2014-hte,silaev2015-lrs} also imply a coupling of
spin and heat. However, a general description of the implications for
the magnetization dynamics, dynamical heat pumping effects, and the behavior
in the non-linear regime at energies comparable to the superconductor
gap $\Delta$, has been lacking.

\begin{figure}
  \includegraphics{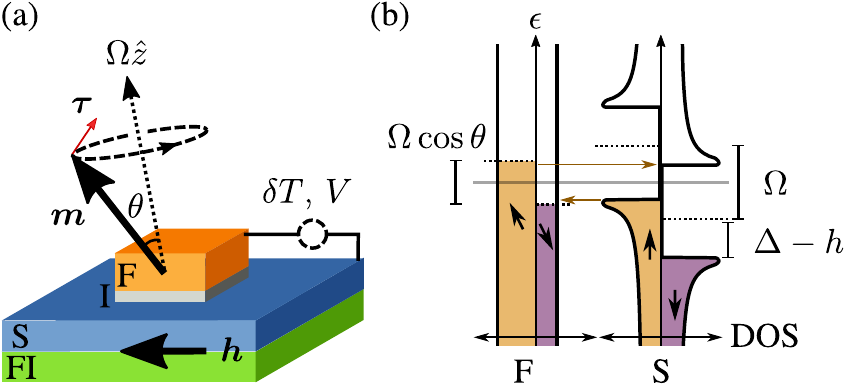}
  \caption{
    \label{fig:scheme}
    (a)
    Schematic ferromagnetic island--superconductor tunnel junction (F/I/S)
    setup.
    The direction $\vec{m}$ of magnetization in F precesses at a rate $\Omega$
    at an angle $\theta$ around the axis ($\hat{z}$) of its effective field.
    Electron tunneling and intrinsic damping produces torque
    $\vec{\tau}$ on $\vec{m}$.
    The superconductor has an internal spin splitting exchange field $\vec{h}$,
    from external magnetic field, or a ferromagnetic insulator (FI) bilayer structure \cite{tedrow1986-spe}. 
    We consider also thermal and electric biasing \((\delta T, V)\).
    (b)
    ``Semiconductor picture'' for pumping,
    in the frame rotating with \(\vec m\) (for $\vec{h}\parallel\hat{z}$).
    Grey solid line is the chemical potential when \(\Omega=0\). Increasing the precession frequency to $\Omega\ne0$
    shifts both the spectrum and the chemical potentials (dashed lines) by \(\Omega\cos\theta\) in F and by \(\Omega\) in S. The exchange field \(\vec{h}\) only shifts the spectrum in S.
  }
\end{figure}

In this work, we fill this gap by constructing a theory which provides a combined
description of pumped charge and heat currents, spin torques, magnetization damping, voltage and thermal bias. We consider a metallic nanomagnet F with a
magnetization precessing at a rate $\Omega$ which is determined by an external magnetic field, the shape of the magnet and the crystal anisotropy, \cite{kittel1948theory} at a slowly varying
angle $\theta$ to the precession axis [Fig.~\ref{fig:scheme}(a)]. The
magnet is tunnel coupled to a superconducting electrode S that also
contains a constant spin-splitting (exchange or Zeeman) field \cite{tedrow1986-spe,tokuyasu1988-pef}.

Main features of the problem can be
understood in a tunneling model, shown schematically in
Fig.~\ref{fig:scheme}(b). Both the spin splitting $h$ and nonzero $\Omega$ shift the
spectrum, whereas $\Omega$ generates also effective spin-dependent
chemical potential shifts \cite{tserkovnyak2005-nmd} providing a
driving force which pumps the currents across the interface. The interplay
of the two enables a coupling between the magnetization dynamics and
the linear-response thermoelectric effect \cite{machon2013-nte,ozaeta2014-hte,bergeret2018-cne}
originating from the spin-selective breaking of the electron-hole symmetry in the superconductor
with respect to the chemical potential.
As a consequence, a
temperature difference between the two systems leads to a thermal spin
torque, which in a suitable
parameter regime yields an anti-damping sufficient to obtain flipping or stable
precession of the nanomagnet. The Onsager counterpart of the thermal spin torque is
a Peltier-type cooling (or heating) driven by the
precessing magnetization.  In the non-linear response, the precession also pumps a charge current, as already
shown in \cite{trif2013-dme}. We discuss the general picture for the
spin-split superconductor, and, in addition to the thermomagnetic effects, find the Keldysh action
[Eq.~\eqref{eq:sTscalar}] describing the stochastic
properties of the S/F junction. The action
allows identifying thermodynamical constraints, current noises, a
spintronic fluctuation theorem and describes the probability distribution
of the magnetization direction and the spectrum of its oscillations.

The manuscript is structured as follows:
We introduce a simple tunneling model in Sec.~\ref{sec:tunneling-model}
and discuss the tunneling currents in Sec.~\ref{sec:tunneling-currents}.
Implications on magnetization dynamics are considered in Sec.~\ref{sec:magnetization-dynamics},
including thermal transport associated with the ferromagnetic resonance
and physics of spin torque oscillators driven by the thermal effects.
In Sec.~\ref{sec:keldysh-action} we focus on studying the stochastic magnetization
dynamics based on a Keldysh action approach to  the tunneling model, and discuss
probability distributions and linewidths for the oscillators.
We conclude in Sec.~\ref{sec:discussion}. Certain details of derivations
are postponed to the Appendixes.

\section{Tunneling model}
\label{sec:tunneling-model}

The main effects can be understood with a tunneling Hamiltonian description
(below $\hbar=e=k_B=1$),
\begin{gather}
  \label{eq:hamiltonian}
  H = H_S + \hat{R}(t) H_{F} \hat{R}(t)^\dagger + \sum_{jj'\sigma}W_{jj'} e^{-iVt} c^\dagger_{j\sigma}d_{j'\sigma} + \mathrm{h.c.}
  \,,
\end{gather}
where $c_{j\sigma}$ and $d_{j\sigma}$ are the F and S conduction
electron operators and $W$ the tunneling matrix elements
for spin/momentum states $\sigma=\pm$, $\vec{p}_j$, and $V$ is a bias voltage.
The Hamiltonian $H_S$ describes the spin-split superconductor \cite{bergeret2018-cne},
and $H_{F}$ the magnet with magnetization parallel to the $\hat{z}$-direction. 
The magnetization direction
$\vec{m}(t)=(\cos\phi\sin\theta,\sin\phi\sin\theta,\cos\theta)$ is
specified by a spin rotation matrix $\hat{R}(t)c_{j\sigma}\hat{R}(t)^\dagger=\sum_{\sigma'}R_{\sigma\sigma'}(t)c_{j\sigma'}$.
In the frame rotating with $R$ \cite{tserkovnyak2005-nmd,tserkovnyak2008-tbe},
assuming $\vec{m}(t)$ varies adiabatically so that an equilibrium electron distribution is maintained,
the Berry phase $\varphi(t)=\int^t\dd{t'}\dot{\phi}(1-\cos\theta)$ can be absorbed
(c.f. Refs.~\onlinecite{flebus2017-tmm}, \onlinecite{shnirman2015-gqn} and Appendix~\ref{app:green}) to the spin rotation
\begin{equation}
  R = e^{-i\phi(t)\sigma_z/2}e^{-i\theta(t)\sigma_y/2}e^{i\phi(t)\sigma_z/2}e^{-i\varphi(t)\sigma_z/2}
  \,,
\end{equation}
where $\sigma_{x/y/z}$ are the spin matrices. 
Varying $\vec{m}(t)$ results to effective spin-dependent voltages \cite{tserkovnyak2008-tbe} in the tunneling
part. For uniform precession, they are $\Omega_{\sigma\sigma'}=(\sigma-\sigma'\cos\theta)\Omega/2$
(see Fig.~\ref{fig:scheme}b).  From the model, we can compute in
leading order in $W$ the tunneling charge, energy, and spin currents
($I_c$, $\dot{E}$, $\vec{I}_s$)
via a standard Green function approach (see Ref.~\onlinecite{bergeret12} and Appendix~\ref{app:tunn}).
The assumption of local equilibrium implies that the rates of
tunneling and other nonequilibrium-generating processes on the
magnetic island should be small compared to electron relaxation.
\cite{ludwig2017-sne,ludwig2019-tds,ludwig2019-cng}

Consider precession with frequency \(\Omega\) around the $z$-axis, $\phi(t)=\Omega{}t$ with
$|\dot{\theta}|\ll\Omega$.
From the above model, we find the time-averaged currents and
$\hbar\overline{\tau_z}=-\overline{(\vec{m}\times{}\vec{I}_s\times\vec{m})_z}$,
\cite{slonczewski1996-cde,tserkovnyak2005-nmd}
the $z$-component of the time-averaged spin transfer torque:
\begin{align}
  \label{eq:avgic}
  \overline{I_c}
  &=
  \frac{G_T}{2e}
  \int_{-\infty}^\infty\dd{\epsilon}
  \sum_{\sigma\sigma'}
  \langle{\sigma}\vert{}\sigma'\rangle^2
  N_{S,\sigma}N_{F,\sigma'}[f_{F} - f_S]
  \,,
  \\
  \label{eq:avgie}
  \overline{\dot E_S}
  &=
  \frac{G_T}{2e^2}
  \int_{-\infty}^\infty\dd{\epsilon}
  \sum_{\sigma\sigma'}
  \epsilon
  \langle{\sigma}\vert{}\sigma'\rangle^2
  N_{S,\sigma}N_{F,\sigma'}[f_{F} - f_S]
  \,,
  \\
  \label{eq:avgtauz}
  \overline{\tau_z}
  &=
  -
  \frac{G_T\sin^2\theta}{8e^2}
  \int_{-\infty}^\infty\dd{\epsilon}\!
  \sum_{\sigma\sigma'}
  \sigma N_{S,\sigma}N_{F,\sigma'}[f_{F} - f_S]
  \,.
\end{align}
Here, $f_F=f_0(\epsilon-V-\Omega_{\sigma\sigma'},T_F)$, $f_S=f_0(\epsilon,T_S)$ are the Fermi
distribution functions in F and S,
$\langle{\sigma}\vert{}\sigma'\rangle^2=(1 + \sigma\sigma'\cos\theta)/2$ the spin overlap between $\vec{m}$ and the $z$-axis,
and $N_{S/F,\sigma=\pm}$
the densities of states (DOS) for up/down spins (quantization axis $\vec{m}(t)$ for F, and $\hat{z}$ for S)
normalized by the Fermi level DOS per spin, and $G_T$ the tunneling conductance.
Of these, Eq.~\eqref{eq:avgic} was previously discussed in Ref.~\onlinecite{trif2013-dme} for $\vec{h} = 0$. 
Using a basic model for F and S, we have $N_{F,\sigma}=1+\sigma{}P$ and $N_{S,\sigma}=\sum_{\pm}\frac{1\pm{}\sigma\hat{h}\cdot\hat{z}}{2}N_0(\epsilon\mp{}h)$,
where $P=(\nu_{F,+}-\nu_{F,-})/(\nu_{F,+}+\nu_{F,-})$ is the spin polarization in terms of the majority/minority Fermi level DOS $\nu_{F,\pm}$, and $N_0(\epsilon)$ the Bardeen-Cooper-Schrieffer density of states \cite{tinkham2004introduction}.
The tunneling described by Eqs.~(\ref{eq:avgic}--\ref{eq:avgtauz}) can be understood in a semiconductor picture, as shown in Fig.~\ref{fig:scheme}b. The broken electron-hole symmetry around the chemical potentials for both spins in S and spin polarization in F results to thermally driven spin currents causing torques, and the rotation-induced potential shifts pump charge and heat currents.

\section{Tunneling currents}
\label{sec:tunneling-currents}

Expanding for small voltage bias $V$,
temperature difference $\delta T=T_S-T_F$, and the precession speed
$\Omega$, the time-averaged currents are described by a linear-response matrix:
\begin{align}
  \label{eq:onsager}
  \begin{pmatrix}
    \overline{I_c}
    \\
    \overline{\dot E_S}
    \\
    \overline{\tau_z}
  \end{pmatrix}
  =
  \begin{pmatrix}
    G & P\alpha\cos\theta & 0
    \\
    P\alpha\cos\theta & G_{\rm th}T & \frac{\alpha}{2}\sin^2\theta
    \\
    0 & -\frac{\alpha}{2}\sin^2\theta & -\frac{G}{4}\sin^2\theta
  \end{pmatrix}
  \begin{pmatrix}
    V
    \\
    -\delta T/T
    \\
    \Omega
  \end{pmatrix}
  \,,
\end{align}
where
$G$ and $G_{\rm th}$ are the linear-response
electrical and thermal conductances.
Here, $\alpha=-(G_T/2)\int_{-\infty}^\infty\dd{\epsilon}\epsilon{}[N_{S,+}(\epsilon)-N_{S,-}(\epsilon)]f_0'(\epsilon)$ is a thermoelectric coefficient, \cite{machon2013-nte,ozaeta2014-hte}
which originates from the exchange field $h$ generating the electron-hole asymmetry in the superconductor.
It is nonzero only when S is both superconducting and has a spin splitting $h\ne0$.
The response matrix $L$ in Eq.~\eqref{eq:onsager} has the Onsager
symmetry $L_{ij}=L_{ji}^{\rm tr}$, where tr refers to time-reversal,
$\alpha^{\rm tr}=-\alpha$, $P^{\rm tr}=-P$.

The coefficient for charge pumping is here zero, unlike in the
ferromagnet--ferromagnet case, \cite{tserkovnyak2008-tbe} 
because the spin-(anti)symmetrized DOS of S is also (anti)symmetric in energy. 
This also suppresses linear-response contributions to charge
current from thermal magnetization fluctuations, \cite{flebus2017-tmm}
which are also related to the magnon spin--Seebeck effect \cite{bauer2012-sc,flebus2017-tmm,kato2019-mts}.

Importantly, the spin splitting of the superconductor enables the
precession to pump energy current at linear response, and as its
Onsager counterpart, there is nonzero thermal spin torque (terms with \(\alpha\neq0\)).  This is
made possible by the nonzero thermoelectric coefficient
\cite{machon2013-nte,ozaeta2014-hte} driving spin currents due to a temperature difference.
This effect is (in metals) parametrically larger by
a factor $\varepsilon_F/\Delta\gg1$ than that from normal-state DOS asymmetry
\cite{hatami2007-tst,bauer2012-sc,ludwig2019-tds} in systems with Fermi energy $\varepsilon_F$.

\subsection{Symmetries}

Let us now consider
the joint
probability $P$ of changes $\delta{}n_s$ and $\delta{}E_S$ in the electron
number and energy of S, and a change $\delta{}m_z$ in the magnetization of F, during
a time interval of length $t_0$.  It satisfies a fluctuation relation
\cite{virtanen2017-spt,utsumi2015-fts}:
\begin{align}
  \notag
  P_{t_0}(\delta{}n,\delta{}E_S,\delta{}m_z)
  &=
  e^{T_F^{-1}V\delta n + (T_S^{-1}-T_F^{-1})\delta E_S + T_F^{-1}\Omega\mathcal{S}\delta{}m_z}
  \\
  &\quad \times P_{t_0}^{\rm tr}(-\delta{}n,-\delta{}E_S,\delta{}m_z)
  \label{eq:fluctuationrelation}
  \,.
\end{align}
Here, we denote $\mathcal{S}={\cal V}M_s/(\hbar\gamma)$ as the
effective macrospin of the ferromagnetic island, $\mathcal{V}$ and
$\gamma$ are
the F volume and gyromagnetic ratio and $M_s$ the
magnetization.  Moreover, $P^{\rm tr}$ corresponds to reversed
polarizations and precession
($N_{S/F,\sigma}\mapsto{}N_{S/F,-\sigma}$, $\Omega\mapsto-\Omega$).
The Onsager symmetry of $L_{ij}$ in Eq.~\eqref{eq:onsager} is a
consequence of fluctuation relations \cite{andrieux04}. The energy
transfer $\delta{}E_F$ into the ferromagnet (generally,
$\delta{}E_F\ne\delta{}E_S$) is determined by energy conservation
$\delta{}E_F+\delta{}E_S=V\delta{}n+\Omega\mathcal{S}\delta{}m_z$,
which implies
$\overline{\dot{E}_S+\dot{E}_F}=\overline{I_c}V-\Omega\overline{\tau_z}$.
These results arise from the symmetries of
Eqs.~(\ref{eq:S0},~\ref{eq:sTscalar}) below, for the case where there
is no external magnetic drive.

\begin{figure}
  \centering
  \includegraphics[width=\columnwidth]{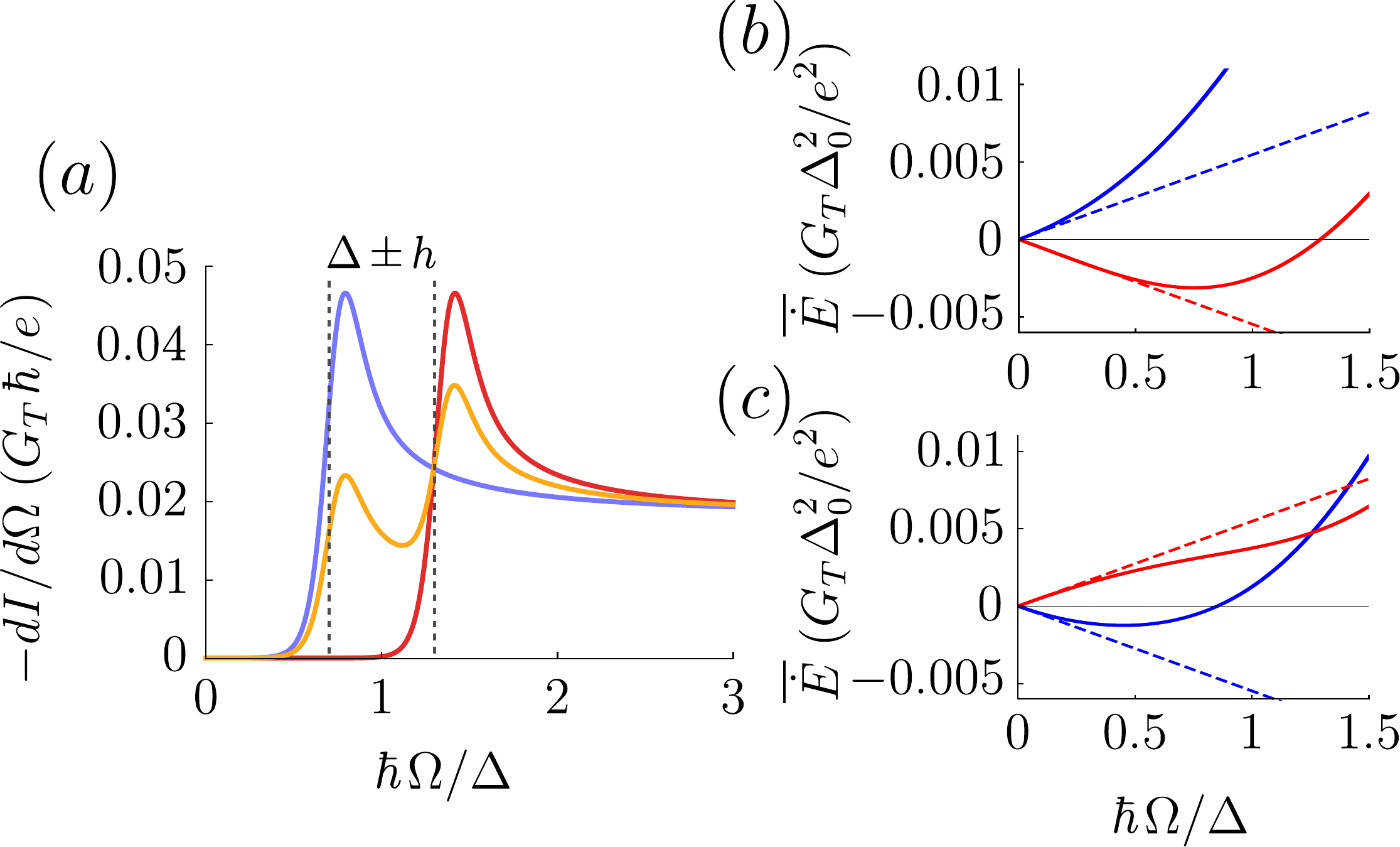}
  \caption{
    (a) Pumped differential current for $T_S = T_F = 0.1\,T_{C}$ where \(T_{C}\) is the critical temperature of the superconductor. Blue, yellow and red lines are for $\vec h =-h \hat z, h \hat x, h \hat z$, respectively. (b) and (c) Energy current into the superconductor $\overline{\dot E_S}$ (blue line) and into the magnet $\overline{\dot E_F}$ (red line) for (b) $\vec h = -h \hat z$ and for (c) $\vec h = h \hat z$. F and S are at temperature $T=0.6\,T_{C}$. Dashed lines represent the linear response. In
    all figures, \(V=0\), $\theta=\frac{\pi}{8}$, $P=1$ and $h=0.3\Delta_0$, where \(\Delta_0\) is the superconductor gap at zero temperature.
  }
  \label{fig:pumping}
\end{figure}

\subsection{Non-linear response}

The pumped charge current is shown in Fig.~\ref{fig:pumping}(a),
and the energy current into S in Fig.~\ref{fig:pumping}(b).
The charge pumping is nonzero above the quasiparticle gap, $|\Omega|\gtrsim\Delta\pm{}h$.
\cite{trif2013-dme} The heat current shows the presence of a region
of cooling of either of the two leads, depending on the relative orientation
of $\vec{h}$ and $\Omega\hat{z}$. Nonzero $h$ enables the N/S cooling effect to
be present already at linear response, similarly as with voltage bias
\cite{bergeret2018-cne,giazotto2006opportunities}.

\begin{figure}
  \centering
  \includegraphics[width=\columnwidth]{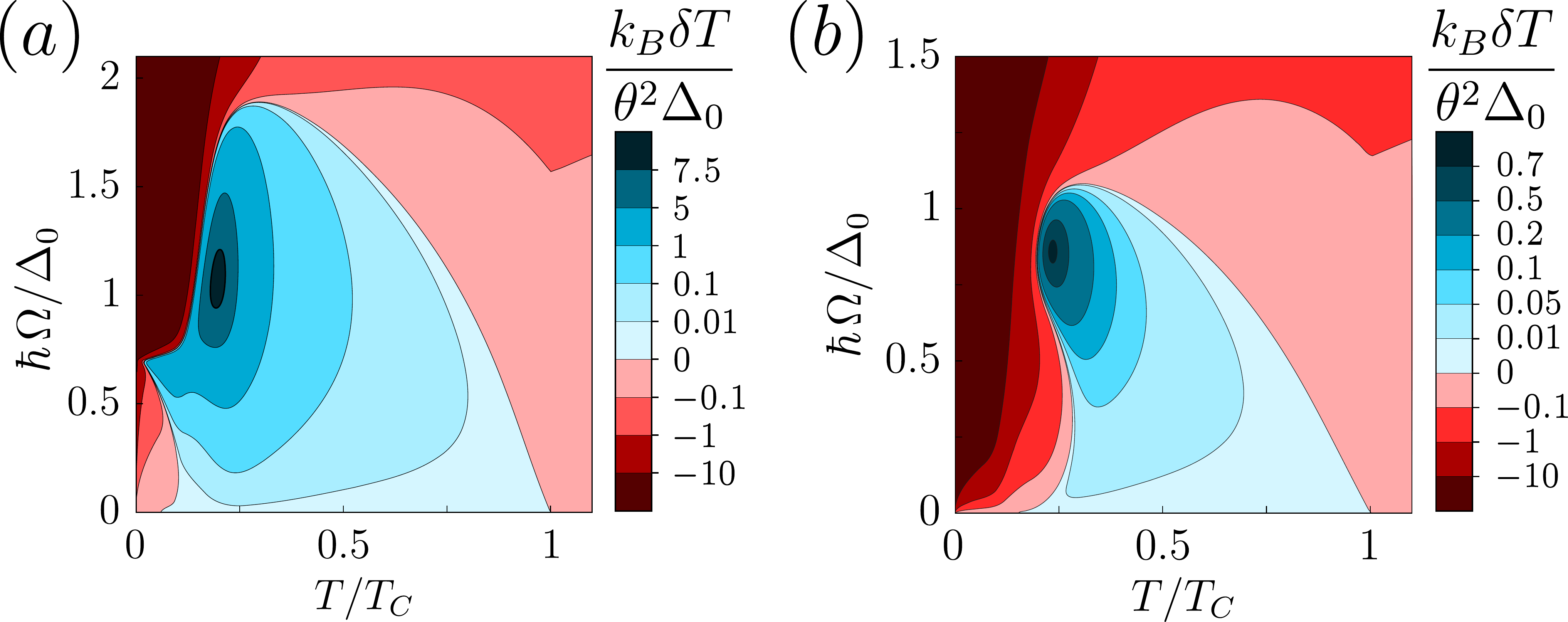}
  \caption{
    Electromagnetically driven FMR induced refrigeration for \(\vec h=-0.3\Delta_0\hat z\), $P=1$, and $A_0=0.1 \hbar G_T/(e^2\mathcal S)$. (a) For $\lambda=0$, and (b) for $\lambda=1$.
    Dynes broadening \(\Gamma=10^{-3}\Delta_0\) was assumed \cite{dynes1984tunneling}. 
  }
  \label{fig:refrigeration}
\end{figure}

\section{Magnetization dynamics}
\label{sec:magnetization-dynamics}

The Landau--Lifshitz--Gilbert--Slonczewski (LLG) equation for
the tilt angle is
\begin{align}
  \label{eq:llg-theta}
  -\mathcal{S}\partial_t\cos\theta = \overline{\tau_z}
  -
  \mathcal{S}A_0\Omega\sin^2\theta
  +
  \eta
  \,,
\end{align}
where the spin transfer torque $\overline{\tau_z}$ is given by
Eq.~\eqref{eq:avgtauz}. 
We include the intrinsic Gilbert damping \cite{tserkovnyak2005-nmd}
phenomenologically, and $A_0$ is the dimensionless damping constant.
Moreover, $\eta$ is a Langevin term describing the torque noise
\cite{chudnovskiy2008-sts,basko2009-sdm,shnirman2015-gqn,virtanen2017-spt}
with the correlation function
$\avg{\eta(t)\eta(t')}=2[D(\theta)+\mathcal{S}A_0T]\sin^2(\theta)\delta(t-t')$;
see below. Equilibrium torques are here included in the LLG effective magnetic field $\Omega\hat{z}$
(see Appendix~\ref{app:tunn}).
We consider the limit of weak damping, where it is sufficient to consider only the equation for the $z$-component.

\subsection{Heat balance in ferromagnetic resonance}
\label{sec:fmr}

Let us consider a ferromagnetic resonance (FMR) \cite{kittel1948theory} in a thin magnetic layer on a spin-split S, driven by a resonant circularly polarized rf magnetic field (at frequency $\omega=\Omega$), and in the case of
S acting as a reservoir at a fixed temperature $T$. The electrical circuit is open, so that no charge flows between F and S. The FMR driving acts as a power source. We assume that a fraction \(\lambda\in[0,1]\) of the power dissipated by the intrinsic Gilbert damping heats the F electrons; the value of $\lambda$ depends on into which bath(s) its microscopic mechanism dissipates the energy (see also Sec.~\ref{sec:intrinsic-damping} below). In a steady state, the total energy current into F, the overall torque, and the charge current are zero:
\begin{align}
  \overline{\dot{E}_{F,\rm tot}}
  =
  \overline{\dot{E}_{F}} + \lambda\overline{P_G}&= 0,\label{eq:heat_balance}\\
    \overline{\tau_{z}} + \overline{\tau_{z,\rm rf}} + \overline{\tau_{z,G}} &= 0,\label{eq:torque_balance}\\
    \overline{I_c} &= 0,\label{eq:charge_balance}
\end{align}
where $\overline{\tau_z}$ and $\overline{I_c}$ are the contributions related to the tunneling between F and S, from Eqs.~(\ref{eq:avgic},\ref{eq:avgtauz}), and $\overline{\dot{E}_{F}}=\overline{I}_cV-\Omega\overline{\tau}_z-\overline{\dot{E}_{S}}$ is found from the tunneling model via a similar calculation as in Eq.~\eqref{eq:avgie}. Moreover, $\overline{\tau_{z,G}}=-\mathcal S A_0 \Omega \sin^2(\theta)$ and $\overline{P_G}=\mathcal{S} A_0 \Omega^2 \sin^2(\theta)$ are the torque due to the intrinsic damping and the rate of work done by it. At resonance, the rf drive creates a torque $\overline{\tau_{z,\rm rf}} = \gamma \mathcal S (\vec m \times \vec h_{\rm rf})_z= \gamma \mathcal{S} h_{\rm rf} \sin\theta$, where $h_{\rm rf}$ is the amplitude of the rf field. From the above it follows that the power
\begin{align}
  \label{eq:absorbed-power-balance}
  \overline{\dot{E}_{S}}+\overline{\dot{E}_{F,\rm tot}}=\overline{P_{\rm rf}} - (1-\lambda)\overline{P_G}
\end{align}
is absorbed by the electron system, where $\overline{P_{\rm rf}} = \Omega \overline{\tau_{z,\rm rf}}$ is the total rf power absorbed at resonance \cite{tserkovnyak2005-nmd}.

Expanding Eqs.~(\ref{eq:avgic}--\ref{eq:avgtauz}) in the linear order in $V$, $\delta T/T$ and $\theta^2$, but not in \(\Omega\), we find the charge and heat currents
\begin{equation}
\begin{pmatrix}
\overline{I_c} \\ \overline{\dot{E}_S} \\ \overline{\tau_z}
\end{pmatrix} = 
\begin{pmatrix}
G & P\alpha & P (G-\widetilde G)\\
P\alpha & G_{\rm th} T & \alpha+\widetilde{\alpha}+\frac{\widetilde G \Omega}{2}\\
0 & 0 & -\widetilde G
\end{pmatrix} 
\begin{pmatrix}
V \\
-\delta T/T \\
\frac{\Omega}{4} \theta^2
\end{pmatrix}.
\end{equation}
Unlike the linear-response matrix in Eq.~\eqref{eq:onsager}, the above matrix is not symmetric, as there is no Onsager reciprocity between \(\overline{\tau_z}\) and \(\theta^2\). The coefficients are
\begin{align}
\widetilde{\alpha} &= \frac{G_T}{2}\int_{-\infty}^{\infty}\mkern-10mu\dd{\epsilon} \sum_\sigma \left(\epsilon-\frac{\sigma\Omega}{2}\right) N_{S,\sigma}(\epsilon) \frac{f_0(\epsilon{-}\sigma\Omega)-f_0(\epsilon)}{\Omega} \\
\widetilde G &= \frac{G_T}{2}\int_{-\infty}^{\infty}\dd{\epsilon} \sum_\sigma \sigma N_{S,\sigma}(\epsilon) \frac{f_0(\epsilon-\sigma\Omega)-f_0(\epsilon)}{\Omega}
\end{align}
These coefficients are defined so that $\lim_{\Omega\to0}\widetilde G = G$ and $\lim_{\Omega\to0}\widetilde \alpha = \alpha$, and they assume the values $\widetilde G_{\rm normal} = G_T$ and $\widetilde\alpha_{\rm normal}= 0$ in the normal state.

The torque balance \eqref{eq:torque_balance} determines the precession angle \(\theta \approx \gamma \mathcal S h_{\rm rf}/{(\mathcal S A_{\rm eff} \Omega})\), where \({\mathcal S A_{\rm eff}} = \mathcal S A_{0}+\frac{\widetilde G}{4}\). To quadratic order in \(h_{\rm rf}\), $\overline{\dot{E}_F} = \widetilde G\Omega^2\theta^2/4-\overline{\dot{E}_S}$. Using this, and the conditions (\ref{eq:heat_balance}) and (\ref{eq:charge_balance}) for heat and charge currents, we find the FMR induced temperature difference and voltage
\begin{equation}
\begin{split}
\begin{pmatrix}
V \\
-\frac{\delta T}{T}
\end{pmatrix}
=
&\begin{pmatrix}
G & P\alpha \\
P\alpha & G_{\rm th} T
\end{pmatrix}^{-1} \times\\
&\begin{pmatrix}
-\frac{P(G-\widetilde G)}{4} \Omega\\
-\frac{\alpha+\widetilde{\alpha}}{4}\Omega + \left[\frac{\widetilde G}{8} + \lambda \mathcal S A_{0}\right]\Omega^2
\end{pmatrix}\theta^2
\end{split}\label{eq:refrigeration}
\end{equation}
The coupling between \(\overline{\dot E_S}\) and \(\theta^2\) is of the linear order in \(\Omega\), whereas the coupling between \(\overline{I_c}\) and \(\theta^2\), the rf power, and the magnetic dissipation are of the quadratic order in \(\Omega\).
Thus, for \(\Omega\ll T\) the induced temperature difference and voltage are
\begin{gather}
    V \simeq \frac{P\alpha}{G T} \delta T \,,\qquad
    \delta T \simeq
    \frac{\alpha
    }{
        2\left(G_{\rm th} - \frac{(P\alpha)^2}{GT}\right)
    }
    \Omega
    \theta^2
    \,.
\end{gather}
The denominator $\widetilde G_{\rm th} = G_{\rm th} - \frac{(P\alpha)^2}{GT}$ is always positive \cite{ozaeta2014-hte}. For \(\Omega\ll T\), F is refrigerated when $\alpha>0$, which corresponds to $\vec h\cdot\hat z<0$. Restoring the SI units, the magnitude of the coefficient between \(\delta T\) and \(\Omega\theta^2\) is $|\hbar\alpha/(\widetilde G_{\rm th}e)| \lesssim \hbar/k_B$
.

\begin{figure}
  \centering
  \includegraphics[width=6cm]{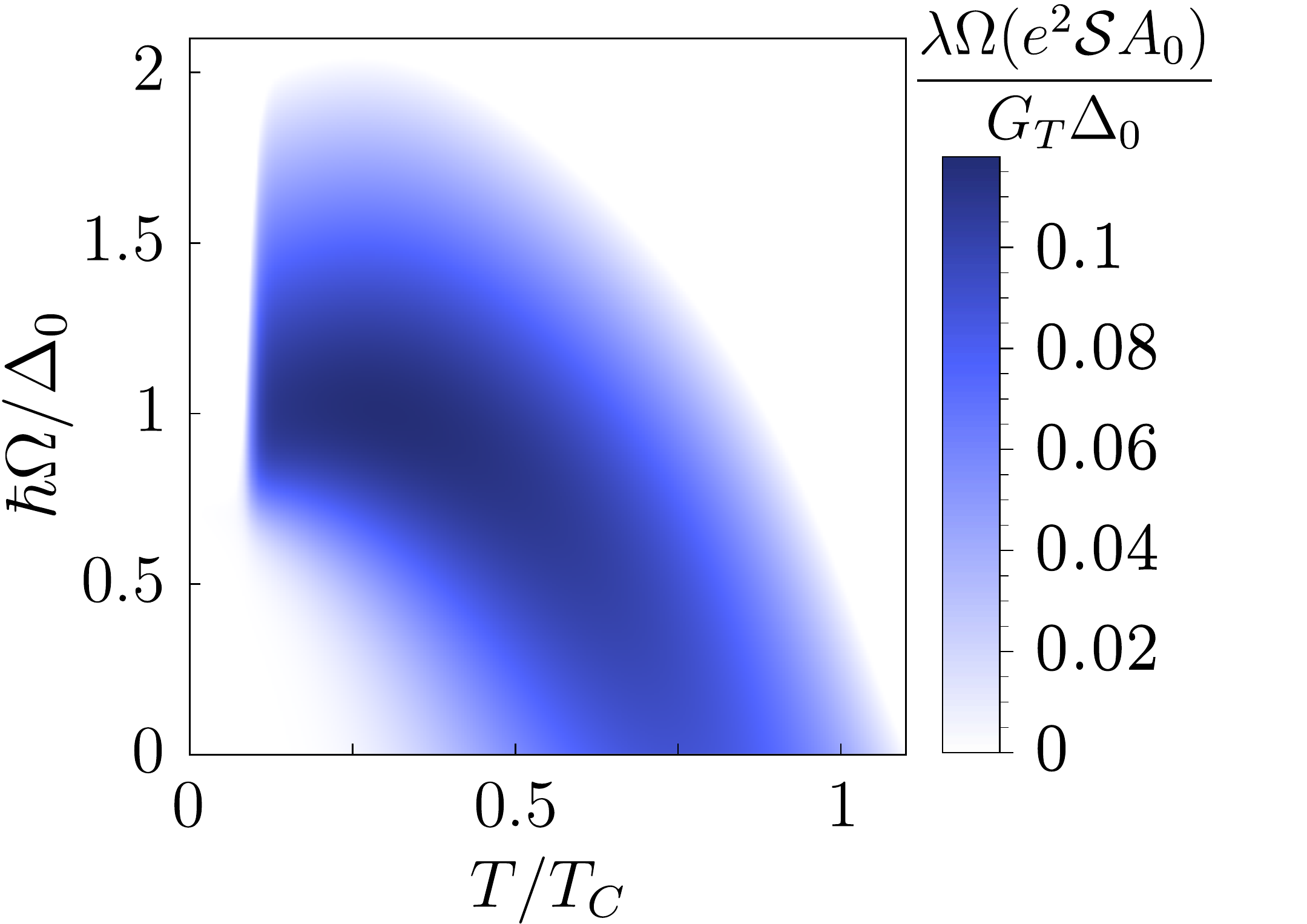}
  \caption{
    Maximum intrinsic damping, expressed as \(\Omega \times \lambda e^2 \mathcal S A_0/(G_T\Delta_0)\), for which the system can be refrigerated, with \(\vec h=-0.3\Delta_0\hat z\), $P=1$ and Dynes broadening \(\Gamma=10^{-5}\Delta_0\). The maximum intrinsic damping is determined by solving $A_0$ from Eq.~\eqref{eq:refrigeration} with \(\delta T=0\).
  }
  \label{fig:maximum_A0}
\end{figure}

At higher frequencies the magnetic dissipation, nonlinearities of \(\widetilde \alpha\) and \(\widetilde G\), and the coupling between charge and precession start to play a role and limit the attainable temperature difference. For $\mathcal S A_0/G_T=0.1$, the magnitude of the effect is illustrated in Fig.~\ref{fig:refrigeration}. The maximum value of \(A_0\) for which refrigeration is possible is shown in Fig.~\ref{fig:maximum_A0} as a function of \(T\) and \(\Omega\). If \(\lambda=1\), the parameter regime is similar to that where the spin-torque driven oscillations occur (see Sec.~\ref{sec:spintorque} below). However, if the intrinsic damping dissipates the energy to systems different from the F conduction electrons ($\lambda <1$), refrigeration is easier to obtain than auto-oscillations. Therefore, measuring the temperature difference $\delta T$ via the thermoelectrically induced voltage $V$ allows for a direct study of the energy dissipation mechanism of the intrinsic Gilbert damping. Note that also in the absence of the spin splitting in S (and therefore $\alpha=0$), it is possible to induce a non-zero voltage via FMR driving \cite{trif2013-dme}. However, that generally requires higher frequencies $\Omega \lesssim \Delta$ than the case analyzed above.

If the thermoelectric coefficient is zero, F always heats up. In the normal state we have
\begin{equation}
    \delta T_{\rm normal} = -\frac{G_T+8\lambda \mathcal S A_{0}}{8G_{\rm th}} \Omega^2\theta^2 < 0,
\end{equation}
which shows the combined heating effect from the different sources of dissipation. However, in that case the induced voltage $V=0$, and the temperature difference would have to be measured via some other mechanism.

\begin{figure}
  \centering
  \includegraphics[width=\columnwidth]{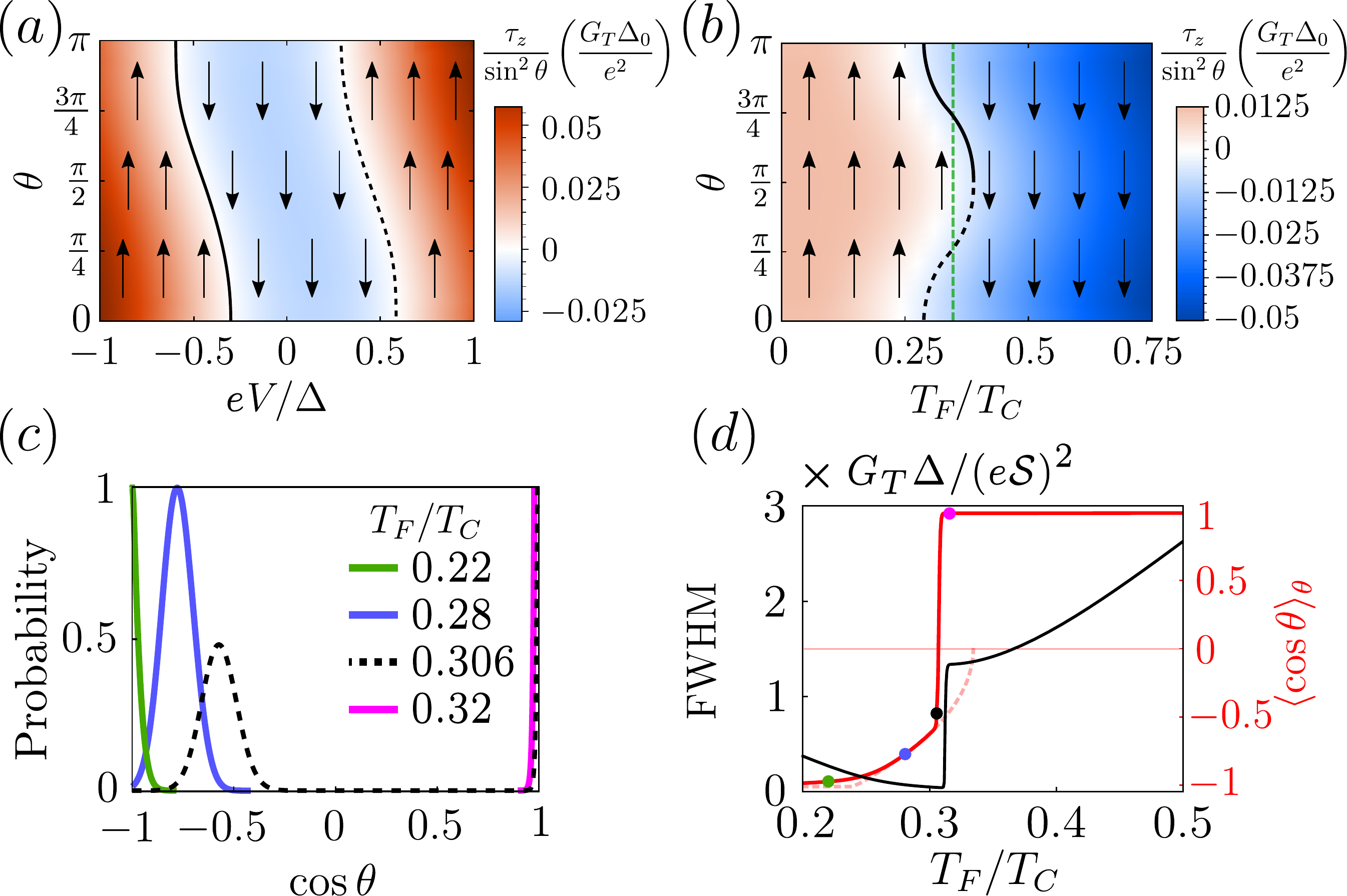}
  \caption{
    (a)
    Torque vs. angle $\theta$ and voltage $V$ at $\Omega=0.3\Delta/\hbar$ for $T_S=T_F=0.5T_{C}$, $\vec h =-0.3\Delta_0 \hat z$ and $P=1$.
    The arrows indicate where the torque drives the angle.
    The solid black line indicates the stable precession angle $\theta_*$, and the dashed line the unstable one. At $V=0$, $\theta_*=0$.
    (b)
    Torque vs. angle and temperature difference at $\Omega = 0.5\Delta/\hbar$ for $T_S=0.5T_C$, $\vec h =0.3\Delta_0 \hat z$, $P=1$, and $V=0$.
    Moreover, $\mathcal{S}A_0=0$. The dashed green line indicates $\delta T_o$.
    (c)
    Magnetization distribution normalized by its maximum value, for a thermally driven spin oscillator with $\mathcal S=100$, $T_S=0.5T_C$, $\vec h =0.3\Delta_0 \hat z$, $P=1$, $V=0$ and $\Omega=0.5\Delta/\hbar$. When $T_F\approx 0.31T_C$ (dashed line), the distribution is significantly bimodal.
    (d)
    Full width at half maximum (FWHM) of the dipole spectrum \(S_{xx}(\omega)\) (black line) and the average magnetization (red line) with $G_T=e^2/\hbar$. The dashed line indicates $\theta_*$ and the dots correspond to panel~(c).
  }
  \label{fig:spintorque}
\end{figure}

\subsection{Spin torques}
\label{sec:spintorque}

The junction also exhibits a \emph{voltage-driven spin torque}.
With an exchange field such that $\vec h\cdot\hat z<0$ and \(\Omega\lesssim 2h\), 
the torque due to tunneling becomes antidamping at large voltages. When it exceeds the intrinsic damping,
the $\theta=0$ equilibrium configuration is destabilized, and a new stable steady-state configuration $\overline{\tau_{z,\rm tot}(\theta_*)}=0$ 
is established.
An example of the signs of the torque and the resulting configuration is
shown in Fig.~\ref{fig:spintorque}(a): The stable angle is $\theta_*=0$ at small voltages,
after which there is a voltage range for which $0<\theta_*<\pi$. There, the system realizes a voltage-driven spin
oscillator \cite{kiselev2003-mon,rippard2004-dci}. At large voltages the stable angle is $\theta_*=\pi$,
corresponding to a torque-driven magnetization flip.

Similarly, the {\em thermal torque} is shown in Fig.~\ref{fig:spintorque}(b).  Due to the nonzero
linear-response coupling, it is antisymmetric in small $\delta T$,
in contrast to the voltage-driven torque.  Consequently,
antidamping regions occur for both signs of $\Omega$. In linear response [Eq.~\eqref{eq:onsager}], for temperature
differences satisfying $\sgn(\alpha)\delta{}T<\delta T_o=[1+e^2\mathcal{S}A_0/(\hbar{}G)]P\hbar\Omega/(2e|\mathfrak{s}|)$, the spin
torque drives $\theta\to0$, damping the precession. Here,
$\mathfrak{s}=-P\alpha/(GT)$ is the junction thermopower, which can be
$|\mathfrak{s}|\gtrsim{}k_B/e$. \cite{ozaeta2014-hte}
Above the critical temperature difference $\delta T_o$,
the thermal spin torque drives the system away from $\theta_*=0$ (or $\theta_*=\pi$ for $\Omega<0$).
The stable
precession angle is shown in Fig.~\ref{fig:spintorque}(b): there is a
range of $\delta T$ in which $\theta_*\ne0,\pi$ and the system
exhibits thermally driven \cite{ludwig2019-tds} spin oscillations.

In Fig.~\ref{fig:spintorque}, we neglect the effect of the intrinsic damping $A_0$ on the magnetization oscillations. However, it is the main obstacle in reaching auto-oscillations in FMR devices, and we estimate its effect here.
For the superconducting systems, 
generally the effective bias $|\mathfrak{s}|\delta T$ can be at most $\Delta$. Considering the value $\delta T_o$ given above, this results to a requirement for the resistance--area product of the S/F junction:
$RA\lesssim{}(RA)_{0}=\frac{\hbar\gamma\Delta}{e^2A_0M_sd_F|\Omega|}\approx\unit[10^{-4}]{\Omega\,\mu{}m^2}\times\frac{\unit[1]{T\,nm}\,\Delta}{\mu_0M_sd_FA_0|\hbar\Omega|}$, where $d_F$ is the ferromagnet thickness.  
Meeting
the requirement is likely challenging. Values
$RA\sim\unit[0.1]{\Omega\,\mu{}m^2}$ have been achieved in $\sim(\unit[100]{nm})^2$ lateral size
magnetic junctions \cite{kiselev2003-mon,nagamine2006-ura}. With such $RA$ and
$\mu_0M_sd_F=\unit[5]{T\,nm}$ (e.g. Co layer \cite{kiselev2003-mon}) and $A_0=0.01$ \cite{tserkovnyak2005-nmd},
the condition is satisfied for $f=|\Omega|/(2\pi)<0.02\Delta/h\approx\unit[1]{GHz}$ (for Al as superconductor).
The FMR refrigeration has a similar requirement but with $\Delta\mapsto\Delta/\lambda$, and hence may
be easier to achieve, if the microscopic mechanism is such that $\lambda<1$.

\section{Keldysh action}
\label{sec:keldysh-action}

To properly describe the metastable states in the magnetization precession,
we need to extend the formalism. The dynamics
beyond average values can be described by an effective action $S=S_0+S_T$ for the
spin including the tunneling, derived
\cite{basko2009-sdm,fransson2008-sdt,zhu2004-nsd,shnirman2015-gqn,virtanen2017-spt,ludwig2017-sne,ludwig2019-tds,ludwig2019-cng} by retaining the Keldysh
structure \cite{kamenevbook} for the orientation of the magnetization mean field.
The action $S$
describes the generating function of the joint probability
distribution $P_{\rm t_0}(\delta n,\delta E_S,\delta E_F,\delta m_z)$
[see Eq.~\eqref{eq:fluctuationrelation}], with a source field
$\chi$, $\xi_S$, $\xi_F$, $\zeta$ associated with each of the arguments.  The
free part reads
\begin{align}
  \label{eq:S0}
  S_0 = 2\mathcal{S}\int_{-\infty}^\infty\mkern-10mu\dd{t}\left[\left(\frac{\zeta}{2}+\phi^q\right)\partial_t(\cos\theta)^c - (\cos\theta)^q(\dot{\phi}^c - \Omega)\right]
  \,,
\end{align}
where $c$ and $q$ denote the symmetric/antisymmetric combinations $x^{c/q}=\frac{x_+\pm{}x_-}{2}$ 
of quantities on the two Keldysh branches ($+/-$), for example
$(\cos\theta)^{c/q}=\frac{1}{2}[\cos(\theta^c+\theta^q)\pm\cos(\theta^c-\theta^q)]$.
Concentrating on slow perturbations
around the semiclassical ($\mathcal{S}\gg1$) precession trajectory $\phi^c(t)=\Omega t$, the tunnelling
action can be expressed as $S_T\simeq{}-i\int_{-\infty}^\infty\dd{t}s_T$ with
\cite{virtanen2017-spt}
\begin{align}
  \label{eq:sTscalar}
  s_T
  &=
  \frac{G_T}{2}
  \int_{-\infty}^\infty\dd{\epsilon}
  \sum_{\sigma\sigma'=\pm}
  N_{F,\sigma'}N_{S,\sigma}
  \bigl\{
  \frac{\cos\theta^q + \sigma\sigma'\cos\theta^c}{2}
  \\\notag&\quad
  \times
  [e^{i\eta_{\sigma\sigma'}}f_F(1-f_S) + e^{-i\eta_{\sigma\sigma'}}f_S(1-f_F)]
  \\\notag&\quad
  -
  \frac{1 + \sigma\sigma'(\cos\theta)^c}{2}[f_F(1-f_S) + f_S(1-f_F)] 
  \bigr\}
  \,,
\end{align}
where $\eta_{\sigma\sigma'}=\chi + \epsilon\xi_S - (\epsilon-V-\Omega_{\sigma\sigma'})\xi_F -
2\phi^q\frac{\Omega_{\sigma\sigma'}}{\Omega}$.  Here, we have neglected
terms that renormalize $\Omega$. For computing time averages, the source fields are taken nonzero
between $t=0$ and $t=t_0$, e.g. $\chi(t)=\chi \theta(|t_0|-|t|)\theta(t\sgn t_0)$.
The results (\ref{eq:avgic}--\ref{eq:avgtauz}) can be found as
$\overline{I_c}=-i\partial_{\chi}s_T\rvert_0$,
$\overline{\dot{E}_S}=-i\partial_{\xi_S}s_T\rvert_0$, and
$\overline{\tau_z}=\frac{1}{2i}\partial_{\phi^q}s_T\rvert_{0}$,
where $\rvert_{0}$ indicates $\phi^q=\theta^q=\chi=\xi_{S/F}=0$.
Expansion around the saddle point gives Eq.~\eqref{eq:llg-theta}, and
the correlator characterizing the spin torque noise is
$D=-\frac{1}{8}\partial_{\phi^q}^2s_T\rvert_{0}\csc^2\theta=-\frac{1}{8}\partial_{\theta^q}^2s_T\rvert_{0}$.

\subsection{Intrinsic damping}
\label{sec:intrinsic-damping}

We can include the phenomenological Gilbert damping term
$A_0\vec{m}\times\dot{\vec{m}}$ of the LLG equation into a
corresponding term in the action, $iS_G=\int_{-\infty}^\infty\dd{t}s_G(t)$. With the weak-damping
assumptions $\dot{\phi}^c\simeq\Omega$, $|\dot{\theta}^c|\ll|\dot{\phi}|$, the leading term
in the torque is produced by $s_G\simeq-2i\mathcal{S}A_0\Omega\sin^2(\theta^c)\phi^q$.

Further reasoning is required for thermodynamic consistency.  Let us first
assume that the Gilbert damping is caused by a coupling that ultimately
dissipates energy into the bath of conduction electrons in F ($\lambda =1$).  We can
express the conservation of energy in conversion of magnetic energy to
energy of conduction electrons as the symmetry $s_G[\xi_F+x,\phi^q+\Omega
  x/2]=s_G[\xi_F,\phi^q]$ for all $x$.  In addition, to preserve the
thermodynamic fluctuation relations and the second law at
equilibrium, the fluctuation symmetry
$s_G[\xi_F,\phi^q]=s_G[iT_F^{-1}-\xi_F,-\phi^q]$ should be fulfilled.
\cite{virtanen2017-spt} The above fixes the series
expansion in $\xi_F$, $\phi^q$, $T_F^{-1}$ to have the form
\begin{align}
  \label{eq:gilbert-action}
  s_G[\xi_F,\phi^q]
  &\simeq
  -2A_0 \mathcal{S} \sin^2(\theta^c)\Bigg[i\Omega\left(\phi^q - \frac{\Omega}{2}\xi_F\right)
  \\\notag&\qquad
  + 2T_F(\phi^q - \frac{\Omega}{2}\xi_F)^2\Bigg] + \ldots
  \,.
\end{align}
If the Gilbert damping dissipates energy directly to multiple baths
(e.g. magnons, phonons), more terms of this form appear, where $\xi_F$
and $T_F$ should be replaced by the corresponding bath variables, and
only a fraction $0\le\lambda\le1$ of the total $A_0$ comes from
conduction electrons.  Including Eq.~\eqref{eq:gilbert-action} in the
total action $S=S_0+S_T+S_G$ then produces e.g. the correlation
function of the Langevin noise terms in Eq.~\eqref{eq:llg-theta}, and
the additional term in the heat balance equation
Eq.~\eqref{eq:heat_balance}.  These are of course possible to find
also directly, by assuming the fluctuation-dissipation theorem, and
reasoning about magnetic work done by the damping.

For the external rf drive, we similarly have a term
$s_{\rm rf}=2i\mathcal{S}\overline{\vec{m}^q\cdot\gamma\vec{h}_{\rm
    rf}}\simeq2i\gamma{}h_{\rm rf}\mathcal{S}\sin(\theta^c)\phi^q$, at
resonance.  It does not obey the above energy conservation symmetry, as
power is externally provided and the mechanism generating
$h_{\rm rf}$ is not included in the model. As a consequence, as noted in
Eq.~\eqref{eq:absorbed-power-balance}
$\overline{\dot{E}_{S,\rm tot}}+\overline{\dot{E}_{F,\rm tot}}\ne0$,
and the fluctuation relation~\eqref{eq:fluctuationrelation} is modified.

\subsection{Spin oscillator}
\label{sec:spin-oscillator}

The probability distribution of the magnetization angle $\theta$ can
be obtained from Eqs.~(\ref{eq:S0},\ref{eq:sTscalar})
\cite{chudnovskiy2008-sts,virtanen2017-spt}, within a
semiclassical method applied to
$\tilde{s}_T=s_T\rvert_{\theta^q=\chi=\xi_j=0}$
\cite{virtanen2017-spt,kamenevbook}.
In this approach, at equilibrium, the fluctuation symmetry
$\tilde{s}_T(\phi^q=-i\Omega/2T)=0$ results to the Boltzmann
distribution $P(\cos\theta)=Ne^{\mathcal{S}\cos(\theta)\Omega/T}$. In
the nonequilibrium driven state ($V\ne0$, $\delta{}T\ne0$), the
distribution deviates from this.

The probability distribution is shown in
Fig.~\ref{fig:spintorque}(c) for the thermally driven
oscillator. The figure shows the spin torque-driven transition from
the magnetization pointing in the direction of the magnetic field ($\cos
\theta=1$) for high $T_F$, to the opposite
direction of the field ($\cos \theta=-1$) at low $T_F$. In the
intermediate range $T_F\approx0.25$--$0.3T_c$, the probability
distribution becomes bimodal, reflecting the two locally stable
configurations in Fig.~\ref{fig:spintorque}(b): one of these
corresponds to the oscillating state.

\subsection{Emission spectrum}
\label{sec:emission-spectrum}

A driven spin oscillator produces electromagnetic emission which can
be detected.  \cite{kiselev2003-mon,rippard2004-dci} This can be
characterized with the classical correlator of the magnetic dipole,
whose spectrum is approximately a Lorentzian centered at frequency
$\Omega$.
The classical spectrum of the magnetic dipole correlator can be
written as
\begin{align}
  S_{xx}(\omega)=\mathcal{S}^2\int_{-\infty}^\infty\dd{t_0}e^{i\omega t_0}\langle{m_x(t_0)m_x(0)}\rangle
  \,,
\end{align}
where $m_x=\cos\phi\sin\theta$, and the average is over the driven steady state
of the system.  To evaluate it, the average over
$\phi$ can be taken first, noting that
$\avg{\cos\phi(t_0)\cos\phi(0)}_\phi=\frac{1}{2}\Re\avg{e^{i\phi(t_0)-i\phi(0)}}_\phi=
\frac{1}{2}\Re\int\DD{\phi^c,\theta^q}e^{iS}e^{i\phi^c(t_0)-i\phi^c(0)}=\frac{1}{2}\Re\int\DD{\phi^c,\theta^q}e^{iS'}$,
where the exponential factor is removed by a shift
$(\cos\theta)^q\mapsto(\cos\theta)^q+\sgn(t_0)\theta(|t_0|-|t|)\theta(t\sgn
t_0)/(2\mathcal{S})$.  For $\mathcal{S}\gg1$, this results to
$S'-S\simeq\Omega{}t_0+i|t_0|\mathcal{S}^{-2}D\csc^2\theta^{c}=:\psi(t_0)$
so that
$\langle{m_x(t_0)m_x(0)}\rangle_\phi\simeq\frac{1}{2}\sin^2\theta{}\Re{}e^{i\psi(t_0)}$.
Evaluating the Fourier transform, we get
\begin{align}
  S_{xx}(\omega)\simeq\frac{1}{2}\sum_\pm\langle{D/[(\omega\pm\Omega)^2+(\mathcal{S}^{-2}D\csc^2\theta^c)^2]}\rangle_\theta
  \,.
\end{align}
A similar calculation is done in Ref.~\onlinecite{chudnovskiy2008-sts}, via
Langevin and Fokker--Planck approaches. 
The remaining average is over the
steady state distribution $P(\cos\theta)$.

The linewidth
of the spectrum [black line in Fig.~\ref{fig:spintorque}(d)] in this nonequilibrium
system is a non-trivial function of the system parameters. 
For $T_F \approx 0.31 T_C$ precession at $\theta_*$ becomes
possible, and as a result the linewidth ($\propto\csc^2\theta$) narrows rapidly,
becoming significantly smaller than the near-equilibrium fluctuations at $\theta \sim 0,\pi$.

\section{Discussion}
\label{sec:discussion}

In this work, we explain how the
thermomagnetoelectric effect of a spin-split superconductor couples
the magnetization in a magnetic tunnel junction to the temperature
difference across it. The thermoelectric coefficient in the
superconducting state is generally large, and enables a magnetic
Peltier effect and thermal spin torque, with prospects for generating
thermally driven oscillations detectable via spectroscopy.
Superconductivity also offers possibilities to characterize
and control the thermal physics via both the electric and magnetic
responses or external field coupling of the magnetization.

\begin{acknowledgments}
We thank A. Di Bernardo for discussions. This work was supported by the Academy of Finland project number 317118, the European Union Horizon 2020 research and innovation programme under grant agreement No.~800923 (SUPERTED), and Jenny and Antti Wihuri Foundation.
\end{acknowledgments}

\bibliography{sfrefs}

\begin{thebibliography}{52}%
\makeatletter
\providecommand \@ifxundefined [1]{%
 \@ifx{#1\undefined}
}%
\providecommand \@ifnum [1]{%
 \ifnum #1\expandafter \@firstoftwo
 \else \expandafter \@secondoftwo
 \fi
}%
\providecommand \@ifx [1]{%
 \ifx #1\expandafter \@firstoftwo
 \else \expandafter \@secondoftwo
 \fi
}%
\providecommand \natexlab [1]{#1}%
\providecommand \enquote  [1]{``#1''}%
\providecommand \bibnamefont  [1]{#1}%
\providecommand \bibfnamefont [1]{#1}%
\providecommand \citenamefont [1]{#1}%
\providecommand \href@noop [0]{\@secondoftwo}%
\providecommand \href [0]{\begingroup \@sanitize@url \@href}%
\providecommand \@href[1]{\@@startlink{#1}\@@href}%
\providecommand \@@href[1]{\endgroup#1\@@endlink}%
\providecommand \@sanitize@url [0]{\catcode `\\12\catcode `\$12\catcode
  `\&12\catcode `\#12\catcode `\^12\catcode `\_12\catcode `\%12\relax}%
\providecommand \@@startlink[1]{}%
\providecommand \@@endlink[0]{}%
\providecommand \url  [0]{\begingroup\@sanitize@url \@url }%
\providecommand \@url [1]{\endgroup\@href {#1}{\urlprefix }}%
\providecommand \urlprefix  [0]{URL }%
\providecommand \Eprint [0]{\href }%
\providecommand \doibase [0]{http://dx.doi.org/}%
\providecommand \selectlanguage [0]{\@gobble}%
\providecommand \bibinfo  [0]{\@secondoftwo}%
\providecommand \bibfield  [0]{\@secondoftwo}%
\providecommand \translation [1]{[#1]}%
\providecommand \BibitemOpen [0]{}%
\providecommand \bibitemStop [0]{}%
\providecommand \bibitemNoStop [0]{.\EOS\space}%
\providecommand \EOS [0]{\spacefactor3000\relax}%
\providecommand \BibitemShut  [1]{\csname bibitem#1\endcsname}%
\let\auto@bib@innerbib\@empty
\bibitem [{\citenamefont {Slonczewski}(1996)}]{slonczewski1996-cde}%
  \BibitemOpen
  \bibfield  {author} {\bibinfo {author} {\bibfnamefont {J.~C.}\ \bibnamefont
  {Slonczewski}},\ }\href {\doibase 10.1016/0304-8853(96)00062-5} {\bibfield
  {journal} {\bibinfo  {journal} {J. Magn. Magn. Mater.}\ }\textbf {\bibinfo
  {volume} {159}},\  (\bibinfo {year} {1996})}\BibitemShut {NoStop}%
\bibitem [{\citenamefont {Johnson}\ and\ \citenamefont
  {Silsbee}(1987)}]{johnson1987-tai}%
  \BibitemOpen
  \bibfield  {author} {\bibinfo {author} {\bibfnamefont {M.}~\bibnamefont
  {Johnson}}\ and\ \bibinfo {author} {\bibfnamefont {R.~H.}\ \bibnamefont
  {Silsbee}},\ }\href {\doibase 10.1103/PhysRevB.35.4959} {\bibfield  {journal}
  {\bibinfo  {journal} {Phys. Rev. B}\ }\textbf {\bibinfo {volume} {35}},\
  \bibinfo {pages} {4959} (\bibinfo {year} {1987})}\BibitemShut {NoStop}%
\bibitem [{\citenamefont {Bauer}\ \emph {et~al.}(2012)\citenamefont {Bauer},
  \citenamefont {Saitoh},\ and\ \citenamefont {van Wees}}]{bauer2012-sc}%
  \BibitemOpen
  \bibfield  {author} {\bibinfo {author} {\bibfnamefont {G.~E.~W.}\
  \bibnamefont {Bauer}}, \bibinfo {author} {\bibfnamefont {E.}~\bibnamefont
  {Saitoh}}, \ and\ \bibinfo {author} {\bibfnamefont {B.~J.}\ \bibnamefont {van
  Wees}},\ }\href {\doibase 10.1038/nmat3301} {\bibfield  {journal} {\bibinfo
  {journal} {Nat. Mater.}\ }\textbf {\bibinfo {volume} {11}},\ \bibinfo {pages}
  {391} (\bibinfo {year} {2012})}\BibitemShut {NoStop}%
\bibitem [{\citenamefont {Tserkovnyak}\ \emph {et~al.}(2002)\citenamefont
  {Tserkovnyak}, \citenamefont {Brataas},\ and\ \citenamefont
  {Bauer}}]{tserkovnyak2002-spm}%
  \BibitemOpen
  \bibfield  {author} {\bibinfo {author} {\bibfnamefont {Y.}~\bibnamefont
  {Tserkovnyak}}, \bibinfo {author} {\bibfnamefont {A.}~\bibnamefont
  {Brataas}}, \ and\ \bibinfo {author} {\bibfnamefont {G.~E.~W.}\ \bibnamefont
  {Bauer}},\ }\href {\doibase 10.1103/PhysRevB.66.224403} {\bibfield  {journal}
  {\bibinfo  {journal} {Phys. Rev. B}\ }\textbf {\bibinfo {volume} {66}},\
  \bibinfo {pages} {224403} (\bibinfo {year} {2002})}\BibitemShut {NoStop}%
\bibitem [{\citenamefont {Bell}\ \emph {et~al.}(2008)\citenamefont {Bell},
  \citenamefont {Milikisyants}, \citenamefont {Huber},\ and\ \citenamefont
  {Aarts}}]{bell2008-sds}%
  \BibitemOpen
  \bibfield  {author} {\bibinfo {author} {\bibfnamefont {C.}~\bibnamefont
  {Bell}}, \bibinfo {author} {\bibfnamefont {S.}~\bibnamefont {Milikisyants}},
  \bibinfo {author} {\bibfnamefont {M.}~\bibnamefont {Huber}}, \ and\ \bibinfo
  {author} {\bibfnamefont {J.}~\bibnamefont {Aarts}},\ }\href {\doibase
  10.1103/PhysRevLett.100.047002} {\bibfield  {journal} {\bibinfo  {journal}
  {Phys. Rev. Lett.}\ }\textbf {\bibinfo {volume} {100}},\ \bibinfo {pages}
  {047002} (\bibinfo {year} {2008})}\BibitemShut {NoStop}%
\bibitem [{\citenamefont {Houzet}(2008)}]{houzet2008-fjj}%
  \BibitemOpen
  \bibfield  {author} {\bibinfo {author} {\bibfnamefont {M.}~\bibnamefont
  {Houzet}},\ }\href {\doibase 10.1103/PhysRevLett.101.057009} {\bibfield
  {journal} {\bibinfo  {journal} {Phys. Rev. Lett.}\ }\textbf {\bibinfo
  {volume} {101}},\ \bibinfo {pages} {057009} (\bibinfo {year}
  {2008})}\BibitemShut {NoStop}%
\bibitem [{\citenamefont {Jeon}\ \emph {et~al.}(2018)\citenamefont {Jeon},
  \citenamefont {Ciccarelli}, \citenamefont {Ferguson}, \citenamefont
  {Kurebayashi}, \citenamefont {Cohen}, \citenamefont {Montiel}, \citenamefont
  {Eschrig}, \citenamefont {Robinson},\ and\ \citenamefont
  {Blamire}}]{jeon2018-esp}%
  \BibitemOpen
  \bibfield  {author} {\bibinfo {author} {\bibfnamefont {K.-R.}\ \bibnamefont
  {Jeon}}, \bibinfo {author} {\bibfnamefont {C.}~\bibnamefont {Ciccarelli}},
  \bibinfo {author} {\bibfnamefont {A.~J.}\ \bibnamefont {Ferguson}}, \bibinfo
  {author} {\bibfnamefont {H.}~\bibnamefont {Kurebayashi}}, \bibinfo {author}
  {\bibfnamefont {L.~F.}\ \bibnamefont {Cohen}}, \bibinfo {author}
  {\bibfnamefont {X.}~\bibnamefont {Montiel}}, \bibinfo {author} {\bibfnamefont
  {M.}~\bibnamefont {Eschrig}}, \bibinfo {author} {\bibfnamefont {J.~W.~A.}\
  \bibnamefont {Robinson}}, \ and\ \bibinfo {author} {\bibfnamefont {M.~G.}\
  \bibnamefont {Blamire}},\ }\href {\doibase 10.1038/s41563-018-0058-9}
  {\bibfield  {journal} {\bibinfo  {journal} {Nat. Mater.}\ }\textbf {\bibinfo
  {volume} {17}},\ \bibinfo {pages} {499} (\bibinfo {year} {2018})}\BibitemShut
  {NoStop}%
\bibitem [{\citenamefont {Yao}\ \emph {et~al.}(2018)\citenamefont {Yao},
  \citenamefont {Song}, \citenamefont {Takamura}, \citenamefont {Cascales},
  \citenamefont {Yuan}, \citenamefont {Ma}, \citenamefont {Yun}, \citenamefont
  {Xie}, \citenamefont {Moodera},\ and\ \citenamefont {Han}}]{yao2018-psd}%
  \BibitemOpen
  \bibfield  {author} {\bibinfo {author} {\bibfnamefont {Y.}~\bibnamefont
  {Yao}}, \bibinfo {author} {\bibfnamefont {Q.}~\bibnamefont {Song}}, \bibinfo
  {author} {\bibfnamefont {Y.}~\bibnamefont {Takamura}}, \bibinfo {author}
  {\bibfnamefont {J.~P.}\ \bibnamefont {Cascales}}, \bibinfo {author}
  {\bibfnamefont {W.}~\bibnamefont {Yuan}}, \bibinfo {author} {\bibfnamefont
  {Y.}~\bibnamefont {Ma}}, \bibinfo {author} {\bibfnamefont {Y.}~\bibnamefont
  {Yun}}, \bibinfo {author} {\bibfnamefont {X.~C.}\ \bibnamefont {Xie}},
  \bibinfo {author} {\bibfnamefont {J.~S.}\ \bibnamefont {Moodera}}, \ and\
  \bibinfo {author} {\bibfnamefont {W.}~\bibnamefont {Han}},\ }\href {\doibase
  10.1103/PhysRevB.97.224414} {\bibfield  {journal} {\bibinfo  {journal} {Phys.
  Rev. B}\ }\textbf {\bibinfo {volume} {97}},\ \bibinfo {pages} {224414}
  (\bibinfo {year} {2018})}\BibitemShut {NoStop}%
\bibitem [{\citenamefont {Jeon}\ \emph {et~al.}(2019)\citenamefont {Jeon},
  \citenamefont {Ciccarelli}, \citenamefont {Kurebayashi}, \citenamefont
  {Cohen}, \citenamefont {Montiel}, \citenamefont {Eschrig}, \citenamefont
  {Wagner}, \citenamefont {Komori}, \citenamefont {Srivastava}, \citenamefont
  {Robinson},\ and\ \citenamefont {Blamire}}]{jeon2019-ems}%
  \BibitemOpen
  \bibfield  {author} {\bibinfo {author} {\bibfnamefont {K.-R.}\ \bibnamefont
  {Jeon}}, \bibinfo {author} {\bibfnamefont {C.}~\bibnamefont {Ciccarelli}},
  \bibinfo {author} {\bibfnamefont {H.}~\bibnamefont {Kurebayashi}}, \bibinfo
  {author} {\bibfnamefont {L.~F.}\ \bibnamefont {Cohen}}, \bibinfo {author}
  {\bibfnamefont {X.}~\bibnamefont {Montiel}}, \bibinfo {author} {\bibfnamefont
  {M.}~\bibnamefont {Eschrig}}, \bibinfo {author} {\bibfnamefont
  {T.}~\bibnamefont {Wagner}}, \bibinfo {author} {\bibfnamefont
  {S.}~\bibnamefont {Komori}}, \bibinfo {author} {\bibfnamefont
  {A.}~\bibnamefont {Srivastava}}, \bibinfo {author} {\bibfnamefont {J.~W.~A.}\
  \bibnamefont {Robinson}}, \ and\ \bibinfo {author} {\bibfnamefont {M.~G.}\
  \bibnamefont {Blamire}},\ }\href {\doibase 10.1103/PhysRevApplied.11.014061}
  {\bibfield  {journal} {\bibinfo  {journal} {Phys. Rev. Applied}\ }\textbf
  {\bibinfo {volume} {11}},\ \bibinfo {pages} {014061} (\bibinfo {year}
  {2019})}\BibitemShut {NoStop}%
\bibitem [{\citenamefont {Rogdakis}\ \emph {et~al.}(2019)\citenamefont
  {Rogdakis}, \citenamefont {Sud}, \citenamefont {Amado}, \citenamefont {Lee},
  \citenamefont {McKenzie-Sell}, \citenamefont {Jeon}, \citenamefont {Cubukcu},
  \citenamefont {Blamire}, \citenamefont {Robinson}, \citenamefont {Cohen},\
  and\ \citenamefont {Kurebayashi}}]{rogdakis2019-stp}%
  \BibitemOpen
  \bibfield  {author} {\bibinfo {author} {\bibfnamefont {K.}~\bibnamefont
  {Rogdakis}}, \bibinfo {author} {\bibfnamefont {A.}~\bibnamefont {Sud}},
  \bibinfo {author} {\bibfnamefont {M.}~\bibnamefont {Amado}}, \bibinfo
  {author} {\bibfnamefont {C.~M.}\ \bibnamefont {Lee}}, \bibinfo {author}
  {\bibfnamefont {L.}~\bibnamefont {McKenzie-Sell}}, \bibinfo {author}
  {\bibfnamefont {K.~R.}\ \bibnamefont {Jeon}}, \bibinfo {author}
  {\bibfnamefont {M.}~\bibnamefont {Cubukcu}}, \bibinfo {author} {\bibfnamefont
  {M.~G.}\ \bibnamefont {Blamire}}, \bibinfo {author} {\bibfnamefont
  {J.~W.~A.}\ \bibnamefont {Robinson}}, \bibinfo {author} {\bibfnamefont
  {L.~F.}\ \bibnamefont {Cohen}}, \ and\ \bibinfo {author} {\bibfnamefont
  {H.}~\bibnamefont {Kurebayashi}},\ }\href {\doibase
  10.1103/PhysRevMaterials.3.014406} {\bibfield  {journal} {\bibinfo  {journal}
  {Phys. Rev. Materials}\ }\textbf {\bibinfo {volume} {3}},\ \bibinfo {pages}
  {014406} (\bibinfo {year} {2019})}\BibitemShut {NoStop}%
\bibitem [{\citenamefont {Morten}\ \emph {et~al.}(2008)\citenamefont {Morten},
  \citenamefont {Brataas}, \citenamefont {Bauer}, \citenamefont {Belzig},\ and\
  \citenamefont {Tserkovnyak}}]{morten2008-pea}%
  \BibitemOpen
  \bibfield  {author} {\bibinfo {author} {\bibfnamefont {J.~P.}\ \bibnamefont
  {Morten}}, \bibinfo {author} {\bibfnamefont {A.}~\bibnamefont {Brataas}},
  \bibinfo {author} {\bibfnamefont {G.~E.~W.}\ \bibnamefont {Bauer}}, \bibinfo
  {author} {\bibfnamefont {W.}~\bibnamefont {Belzig}}, \ and\ \bibinfo {author}
  {\bibfnamefont {Y.}~\bibnamefont {Tserkovnyak}},\ }\href {\doibase
  10.1209/0295-5075/84/57008} {\bibfield  {journal} {\bibinfo  {journal} {EPL}\
  }\textbf {\bibinfo {volume} {84}},\ \bibinfo {pages} {57008} (\bibinfo {year}
  {2008})}\BibitemShut {NoStop}%
\bibitem [{\citenamefont {Skadsem}\ \emph {et~al.}(2011)\citenamefont
  {Skadsem}, \citenamefont {Brataas}, \citenamefont {Martinek},\ and\
  \citenamefont {Tserkovnyak}}]{skadsem2011-frv}%
  \BibitemOpen
  \bibfield  {author} {\bibinfo {author} {\bibfnamefont {H.~J.}\ \bibnamefont
  {Skadsem}}, \bibinfo {author} {\bibfnamefont {A.}~\bibnamefont {Brataas}},
  \bibinfo {author} {\bibfnamefont {J.}~\bibnamefont {Martinek}}, \ and\
  \bibinfo {author} {\bibfnamefont {Y.}~\bibnamefont {Tserkovnyak}},\ }\href
  {\doibase 10.1103/PhysRevB.84.104420} {\bibfield  {journal} {\bibinfo
  {journal} {Phys. Rev. B}\ }\textbf {\bibinfo {volume} {84}},\ \bibinfo
  {pages} {104420} (\bibinfo {year} {2011})}\BibitemShut {NoStop}%
\bibitem [{\citenamefont {Inoue}\ \emph {et~al.}(2017)\citenamefont {Inoue},
  \citenamefont {Ichioka},\ and\ \citenamefont {Adachi}}]{adachi2017-spi}%
  \BibitemOpen
  \bibfield  {author} {\bibinfo {author} {\bibfnamefont {M.}~\bibnamefont
  {Inoue}}, \bibinfo {author} {\bibfnamefont {M.}~\bibnamefont {Ichioka}}, \
  and\ \bibinfo {author} {\bibfnamefont {H.}~\bibnamefont {Adachi}},\ }\href
  {\doibase 10.1103/PhysRevB.96.024414} {\bibfield  {journal} {\bibinfo
  {journal} {Phys. Rev. B}\ }\textbf {\bibinfo {volume} {96}},\ \bibinfo
  {pages} {024414} (\bibinfo {year} {2017})}\BibitemShut {NoStop}%
\bibitem [{\citenamefont {Teber}\ \emph {et~al.}(2010)\citenamefont {Teber},
  \citenamefont {Holmqvist},\ and\ \citenamefont
  {Fogelstr{\"{o}}m}}]{teber2010-tmd}%
  \BibitemOpen
  \bibfield  {author} {\bibinfo {author} {\bibfnamefont {S.}~\bibnamefont
  {Teber}}, \bibinfo {author} {\bibfnamefont {C.}~\bibnamefont {Holmqvist}}, \
  and\ \bibinfo {author} {\bibfnamefont {M.}~\bibnamefont {Fogelstr{\"{o}}m}},\
  }\href {\doibase 10.1103/PhysRevB.81.174503} {\bibfield  {journal} {\bibinfo
  {journal} {Phys. Rev. B}\ }\textbf {\bibinfo {volume} {81}},\ \bibinfo
  {pages} {174503} (\bibinfo {year} {2010})}\BibitemShut {NoStop}%
\bibitem [{\citenamefont {Richard}\ \emph {et~al.}(2012)\citenamefont
  {Richard}, \citenamefont {Houzet},\ and\ \citenamefont
  {Meyer}}]{richard2012-aci}%
  \BibitemOpen
  \bibfield  {author} {\bibinfo {author} {\bibfnamefont {C.}~\bibnamefont
  {Richard}}, \bibinfo {author} {\bibfnamefont {M.}~\bibnamefont {Houzet}}, \
  and\ \bibinfo {author} {\bibfnamefont {J.~S.}\ \bibnamefont {Meyer}},\ }\href
  {\doibase 10.1103/PhysRevLett.109.057002} {\bibfield  {journal} {\bibinfo
  {journal} {Phys. Rev. Lett.}\ }\textbf {\bibinfo {volume} {109}},\ \bibinfo
  {pages} {057002} (\bibinfo {year} {2012})}\BibitemShut {NoStop}%
\bibitem [{\citenamefont {Holmqvist}\ \emph {et~al.}(2014)\citenamefont
  {Holmqvist}, \citenamefont {Fogelstr{\"{o}}m},\ and\ \citenamefont
  {Belzig}}]{holmqvist2014-sps}%
  \BibitemOpen
  \bibfield  {author} {\bibinfo {author} {\bibfnamefont {C.}~\bibnamefont
  {Holmqvist}}, \bibinfo {author} {\bibfnamefont {M.}~\bibnamefont
  {Fogelstr{\"{o}}m}}, \ and\ \bibinfo {author} {\bibfnamefont
  {W.}~\bibnamefont {Belzig}},\ }\href {\doibase 10.1103/PhysRevB.90.014516}
  {\bibfield  {journal} {\bibinfo  {journal} {Phys. Rev. B}\ }\textbf {\bibinfo
  {volume} {90}},\ \bibinfo {pages} {014516} (\bibinfo {year}
  {2014})}\BibitemShut {NoStop}%
\bibitem [{\citenamefont {Hammar}\ and\ \citenamefont
  {Fransson}(2017)}]{hammar2017-tsd}%
  \BibitemOpen
  \bibfield  {author} {\bibinfo {author} {\bibfnamefont {H.}~\bibnamefont
  {Hammar}}\ and\ \bibinfo {author} {\bibfnamefont {J.}~\bibnamefont
  {Fransson}},\ }\href {\doibase 10.1103/PhysRevB.96.214401} {\bibfield
  {journal} {\bibinfo  {journal} {Phys. Rev. B}\ }\textbf {\bibinfo {volume}
  {96}},\ \bibinfo {pages} {214401} (\bibinfo {year} {2017})}\BibitemShut
  {NoStop}%
\bibitem [{\citenamefont {Kato}\ \emph {et~al.}(2019)\citenamefont {Kato},
  \citenamefont {Ohnuma}, \citenamefont {Matsuo}, \citenamefont {Rech},
  \citenamefont {Jonckheere},\ and\ \citenamefont {Martin}}]{kato2019-mts}%
  \BibitemOpen
  \bibfield  {author} {\bibinfo {author} {\bibfnamefont {T.}~\bibnamefont
  {Kato}}, \bibinfo {author} {\bibfnamefont {Y.}~\bibnamefont {Ohnuma}},
  \bibinfo {author} {\bibfnamefont {M.}~\bibnamefont {Matsuo}}, \bibinfo
  {author} {\bibfnamefont {J.}~\bibnamefont {Rech}}, \bibinfo {author}
  {\bibfnamefont {T.}~\bibnamefont {Jonckheere}}, \ and\ \bibinfo {author}
  {\bibfnamefont {T.}~\bibnamefont {Martin}},\ }\href {\doibase
  10.1103/PhysRevB.99.144411} {\bibfield  {journal} {\bibinfo  {journal} {Phys.
  Rev. B}\ }\textbf {\bibinfo {volume} {99}},\ \bibinfo {pages} {144411}
  (\bibinfo {year} {2019})}\BibitemShut {NoStop}%
\bibitem [{\citenamefont {Dutta}\ \emph {et~al.}(2017)\citenamefont {Dutta},
  \citenamefont {Saha},\ and\ \citenamefont
  {Jayannavar}}]{dutta2017thermoelectric}%
  \BibitemOpen
  \bibfield  {author} {\bibinfo {author} {\bibfnamefont {P.}~\bibnamefont
  {Dutta}}, \bibinfo {author} {\bibfnamefont {A.}~\bibnamefont {Saha}}, \ and\
  \bibinfo {author} {\bibfnamefont {A.}~\bibnamefont {Jayannavar}},\ }\href
  {\doibase 10.1103/PhysRevB.96.115404} {\bibfield  {journal} {\bibinfo
  {journal} {Phys. Rev. B}\ }\textbf {\bibinfo {volume} {96}},\ \bibinfo
  {pages} {115404} (\bibinfo {year} {2017})}\BibitemShut {NoStop}%
\bibitem [{\citenamefont {Machon}\ \emph {et~al.}(2013)\citenamefont {Machon},
  \citenamefont {Eschrig},\ and\ \citenamefont {Belzig}}]{machon2013-nte}%
  \BibitemOpen
  \bibfield  {author} {\bibinfo {author} {\bibfnamefont {P.}~\bibnamefont
  {Machon}}, \bibinfo {author} {\bibfnamefont {M.}~\bibnamefont {Eschrig}}, \
  and\ \bibinfo {author} {\bibfnamefont {W.}~\bibnamefont {Belzig}},\ }\href
  {\doibase 10.1103/PhysRevLett.110.047002} {\bibfield  {journal} {\bibinfo
  {journal} {Phys. Rev. Lett.}\ }\textbf {\bibinfo {volume} {110}},\ \bibinfo
  {pages} {047002} (\bibinfo {year} {2013})}\BibitemShut {NoStop}%
\bibitem [{\citenamefont {Ozaeta}\ \emph {et~al.}(2014)\citenamefont {Ozaeta},
  \citenamefont {Virtanen}, \citenamefont {Bergeret},\ and\ \citenamefont
  {Heikkil{\"{a}}}}]{ozaeta2014-hte}%
  \BibitemOpen
  \bibfield  {author} {\bibinfo {author} {\bibfnamefont {A.}~\bibnamefont
  {Ozaeta}}, \bibinfo {author} {\bibfnamefont {P.}~\bibnamefont {Virtanen}},
  \bibinfo {author} {\bibfnamefont {F.~S.}\ \bibnamefont {Bergeret}}, \ and\
  \bibinfo {author} {\bibfnamefont {T.~T.}\ \bibnamefont {Heikkil{\"{a}}}},\
  }\href {\doibase 10.1103/PhysRevLett.112.057001} {\bibfield  {journal}
  {\bibinfo  {journal} {Phys. Rev. Lett.}\ }\textbf {\bibinfo {volume} {112}},\
  \bibinfo {pages} {057001} (\bibinfo {year} {2014})}\BibitemShut {NoStop}%
\bibitem [{\citenamefont {Silaev}\ \emph {et~al.}(2015)\citenamefont {Silaev},
  \citenamefont {Virtanen}, \citenamefont {Bergeret},\ and\ \citenamefont
  {Heikkil{\"{a}}}}]{silaev2015-lrs}%
  \BibitemOpen
  \bibfield  {author} {\bibinfo {author} {\bibfnamefont {M.}~\bibnamefont
  {Silaev}}, \bibinfo {author} {\bibfnamefont {P.}~\bibnamefont {Virtanen}},
  \bibinfo {author} {\bibfnamefont {F.~S.}\ \bibnamefont {Bergeret}}, \ and\
  \bibinfo {author} {\bibfnamefont {T.~T.}\ \bibnamefont {Heikkil{\"{a}}}},\
  }\href {\doibase 10.1103/PhysRevLett.114.167002} {\bibfield  {journal}
  {\bibinfo  {journal} {Phys. Rev. Lett.}\ }\textbf {\bibinfo {volume} {114}},\
  \bibinfo {pages} {167002} (\bibinfo {year} {2015})}\BibitemShut {NoStop}%
\bibitem [{\citenamefont {Bergeret}\ \emph {et~al.}(2018)\citenamefont
  {Bergeret}, \citenamefont {Silaev}, \citenamefont {Virtanen},\ and\
  \citenamefont {Heikkil{\"{a}}}}]{bergeret2018-cne}%
  \BibitemOpen
  \bibfield  {author} {\bibinfo {author} {\bibfnamefont {F.~S.}\ \bibnamefont
  {Bergeret}}, \bibinfo {author} {\bibfnamefont {M.}~\bibnamefont {Silaev}},
  \bibinfo {author} {\bibfnamefont {P.}~\bibnamefont {Virtanen}}, \ and\
  \bibinfo {author} {\bibfnamefont {T.~T.}\ \bibnamefont {Heikkil{\"{a}}}},\
  }\href {\doibase 10.1103/RevModPhys.90.041001} {\bibfield  {journal}
  {\bibinfo  {journal} {Rev. Mod. Phys.}\ }\textbf {\bibinfo {volume} {90}},\
  \bibinfo {pages} {041001} (\bibinfo {year} {2018})}\BibitemShut {NoStop}%
\bibitem [{\citenamefont {Heikkil\"a}\ \emph {et~al.}(2019)\citenamefont
  {Heikkil\"a}, \citenamefont {Silaev}, \citenamefont {Virtanen},\ and\
  \citenamefont {Bergeret}}]{heikkila2019-tes}%
  \BibitemOpen
  \bibfield  {author} {\bibinfo {author} {\bibfnamefont {T.~T.}\ \bibnamefont
  {Heikkil\"a}}, \bibinfo {author} {\bibfnamefont {M.}~\bibnamefont {Silaev}},
  \bibinfo {author} {\bibfnamefont {P.}~\bibnamefont {Virtanen}}, \ and\
  \bibinfo {author} {\bibfnamefont {F.~S.}\ \bibnamefont {Bergeret}},\ }\href
  {\doibase 10.1016/j.progsurf.2019.100540} {\bibfield  {journal} {\bibinfo
  {journal} {Prog. Surf. Sci.}\ }\textbf {\bibinfo {volume} {94}},\ \bibinfo
  {pages} {100540} (\bibinfo {year} {2019})}\BibitemShut {NoStop}%
\bibitem [{\citenamefont {Tedrow}\ \emph {et~al.}(1986)\citenamefont {Tedrow},
  \citenamefont {Tkaczyk},\ and\ \citenamefont {Kumar}}]{tedrow1986-spe}%
  \BibitemOpen
  \bibfield  {author} {\bibinfo {author} {\bibfnamefont {P.~M.}\ \bibnamefont
  {Tedrow}}, \bibinfo {author} {\bibfnamefont {J.~E.}\ \bibnamefont {Tkaczyk}},
  \ and\ \bibinfo {author} {\bibfnamefont {A.}~\bibnamefont {Kumar}},\ }\href
  {\doibase 10.1103/PhysRevLett.56.1746} {\bibfield  {journal} {\bibinfo
  {journal} {Phys. Rev. Lett.}\ }\textbf {\bibinfo {volume} {56}},\ \bibinfo
  {pages} {1746} (\bibinfo {year} {1986})}\BibitemShut {NoStop}%
\bibitem [{\citenamefont {Kittel}(1948)}]{kittel1948theory}%
  \BibitemOpen
  \bibfield  {author} {\bibinfo {author} {\bibfnamefont {C.}~\bibnamefont
  {Kittel}},\ }\href {\doibase 10.1103/PhysRev.73.155} {\bibfield  {journal}
  {\bibinfo  {journal} {Phys. Rev.}\ }\textbf {\bibinfo {volume} {73}},\
  \bibinfo {pages} {155} (\bibinfo {year} {1948})}\BibitemShut {NoStop}%
\bibitem [{\citenamefont {Tokuyasu}\ \emph {et~al.}(1988)\citenamefont
  {Tokuyasu}, \citenamefont {Sauls},\ and\ \citenamefont
  {Rainer}}]{tokuyasu1988-pef}%
  \BibitemOpen
  \bibfield  {author} {\bibinfo {author} {\bibfnamefont {T.}~\bibnamefont
  {Tokuyasu}}, \bibinfo {author} {\bibfnamefont {J.~A.}\ \bibnamefont {Sauls}},
  \ and\ \bibinfo {author} {\bibfnamefont {D.}~\bibnamefont {Rainer}},\ }\href
  {\doibase 10.1103/PhysRevB.38.8823} {\bibfield  {journal} {\bibinfo
  {journal} {Phys. Rev. B}\ }\textbf {\bibinfo {volume} {38}},\ \bibinfo
  {pages} {8823} (\bibinfo {year} {1988})}\BibitemShut {NoStop}%
\bibitem [{\citenamefont {Tserkovnyak}\ \emph {et~al.}(2005)\citenamefont
  {Tserkovnyak}, \citenamefont {Brataas}, \citenamefont {Bauer},\ and\
  \citenamefont {Halperin}}]{tserkovnyak2005-nmd}%
  \BibitemOpen
  \bibfield  {author} {\bibinfo {author} {\bibfnamefont {Y.}~\bibnamefont
  {Tserkovnyak}}, \bibinfo {author} {\bibfnamefont {A.}~\bibnamefont
  {Brataas}}, \bibinfo {author} {\bibfnamefont {G.~E.~W.}\ \bibnamefont
  {Bauer}}, \ and\ \bibinfo {author} {\bibfnamefont {B.~I.}\ \bibnamefont
  {Halperin}},\ }\href {\doibase 10.1103/RevModPhys.77.1375} {\bibfield
  {journal} {\bibinfo  {journal} {Rev. Mod. Phys.}\ }\textbf {\bibinfo {volume}
  {77}},\ \bibinfo {pages} {1375} (\bibinfo {year} {2005})}\BibitemShut
  {NoStop}%
\bibitem [{\citenamefont {Trif}\ and\ \citenamefont
  {Tserkovnyak}(2013)}]{trif2013-dme}%
  \BibitemOpen
  \bibfield  {author} {\bibinfo {author} {\bibfnamefont {M.}~\bibnamefont
  {Trif}}\ and\ \bibinfo {author} {\bibfnamefont {Y.}~\bibnamefont
  {Tserkovnyak}},\ }\href {\doibase 10.1103/PhysRevLett.111.087602} {\bibfield
  {journal} {\bibinfo  {journal} {Phys. Rev. Lett.}\ }\textbf {\bibinfo
  {volume} {111}},\ \bibinfo {pages} {087602} (\bibinfo {year}
  {2013})}\BibitemShut {NoStop}%
\bibitem [{\citenamefont {Tserkovnyak}\ \emph {et~al.}(2008)\citenamefont
  {Tserkovnyak}, \citenamefont {Moriyama},\ and\ \citenamefont
  {Xiao}}]{tserkovnyak2008-tbe}%
  \BibitemOpen
  \bibfield  {author} {\bibinfo {author} {\bibfnamefont {Y.}~\bibnamefont
  {Tserkovnyak}}, \bibinfo {author} {\bibfnamefont {T.}~\bibnamefont
  {Moriyama}}, \ and\ \bibinfo {author} {\bibfnamefont {J.~Q.}\ \bibnamefont
  {Xiao}},\ }\href {\doibase 10.1103/PhysRevB.78.020401} {\bibfield  {journal}
  {\bibinfo  {journal} {Phys. Rev. B}\ }\textbf {\bibinfo {volume} {78}},\
  \bibinfo {pages} {020401} (\bibinfo {year} {2008})}\BibitemShut {NoStop}%
\bibitem [{\citenamefont {Flebus}\ \emph {et~al.}(2017)\citenamefont {Flebus},
  \citenamefont {Bauer}, \citenamefont {Duine},\ and\ \citenamefont
  {Tserkovnyak}}]{flebus2017-tmm}%
  \BibitemOpen
  \bibfield  {author} {\bibinfo {author} {\bibfnamefont {B.}~\bibnamefont
  {Flebus}}, \bibinfo {author} {\bibfnamefont {G.~E.~W.}\ \bibnamefont
  {Bauer}}, \bibinfo {author} {\bibfnamefont {R.~A.}\ \bibnamefont {Duine}}, \
  and\ \bibinfo {author} {\bibfnamefont {Y.}~\bibnamefont {Tserkovnyak}},\
  }\href {\doibase 10.1103/PhysRevB.96.094429} {\bibfield  {journal} {\bibinfo
  {journal} {Phys. Rev. B}\ }\textbf {\bibinfo {volume} {96}},\ \bibinfo
  {pages} {094429} (\bibinfo {year} {2017})}\BibitemShut {NoStop}%
\bibitem [{\citenamefont {Shnirman}\ \emph {et~al.}(2015)\citenamefont
  {Shnirman}, \citenamefont {Gefen}, \citenamefont {Saha}, \citenamefont
  {Burmistrov}, \citenamefont {Kiselev},\ and\ \citenamefont
  {Altland}}]{shnirman2015-gqn}%
  \BibitemOpen
  \bibfield  {author} {\bibinfo {author} {\bibfnamefont {A.}~\bibnamefont
  {Shnirman}}, \bibinfo {author} {\bibfnamefont {Y.}~\bibnamefont {Gefen}},
  \bibinfo {author} {\bibfnamefont {A.}~\bibnamefont {Saha}}, \bibinfo {author}
  {\bibfnamefont {I.~S.}\ \bibnamefont {Burmistrov}}, \bibinfo {author}
  {\bibfnamefont {M.~N.}\ \bibnamefont {Kiselev}}, \ and\ \bibinfo {author}
  {\bibfnamefont {A.}~\bibnamefont {Altland}},\ }\href {\doibase
  10.1103/PhysRevLett.114.176806} {\bibfield  {journal} {\bibinfo  {journal}
  {Phys. Rev. Lett.}\ }\textbf {\bibinfo {volume} {114}},\ \bibinfo {pages}
  {176806} (\bibinfo {year} {2015})}\BibitemShut {NoStop}%
\bibitem [{\citenamefont {Bergeret}\ \emph {et~al.}(2012)\citenamefont
  {Bergeret}, \citenamefont {Verso},\ and\ \citenamefont
  {Volkov}}]{bergeret12}%
  \BibitemOpen
  \bibfield  {author} {\bibinfo {author} {\bibfnamefont {F.~S.}\ \bibnamefont
  {Bergeret}}, \bibinfo {author} {\bibfnamefont {A.}~\bibnamefont {Verso}}, \
  and\ \bibinfo {author} {\bibfnamefont {A.~F.}\ \bibnamefont {Volkov}},\
  }\href {\doibase 10.1103/PhysRevB.86.214516} {\bibfield  {journal} {\bibinfo
  {journal} {Phys. Rev. B}\ }\textbf {\bibinfo {volume} {86}},\ \bibinfo
  {pages} {214516} (\bibinfo {year} {2012})}\BibitemShut {NoStop}%
\bibitem [{\citenamefont {Ludwig}\ \emph {et~al.}(2017)\citenamefont {Ludwig},
  \citenamefont {Burmistrov}, \citenamefont {Gefen},\ and\ \citenamefont
  {Shnirman}}]{ludwig2017-sne}%
  \BibitemOpen
  \bibfield  {author} {\bibinfo {author} {\bibfnamefont {T.}~\bibnamefont
  {Ludwig}}, \bibinfo {author} {\bibfnamefont {I.~S.}\ \bibnamefont
  {Burmistrov}}, \bibinfo {author} {\bibfnamefont {Y.}~\bibnamefont {Gefen}}, \
  and\ \bibinfo {author} {\bibfnamefont {A.}~\bibnamefont {Shnirman}},\ }\href
  {\doibase 10.1103/PhysRevB.95.075425} {\bibfield  {journal} {\bibinfo
  {journal} {Phys. Rev. B}\ }\textbf {\bibinfo {volume} {95}},\ \bibinfo
  {pages} {075425} (\bibinfo {year} {2017})}\BibitemShut {NoStop}%
\bibitem [{\citenamefont {Ludwig}\ \emph
  {et~al.}(2019{\natexlab{a}})\citenamefont {Ludwig}, \citenamefont
  {Burmistrov}, \citenamefont {Gefen},\ and\ \citenamefont
  {Shnirman}}]{ludwig2019-tds}%
  \BibitemOpen
  \bibfield  {author} {\bibinfo {author} {\bibfnamefont {T.}~\bibnamefont
  {Ludwig}}, \bibinfo {author} {\bibfnamefont {I.~S.}\ \bibnamefont
  {Burmistrov}}, \bibinfo {author} {\bibfnamefont {Y.}~\bibnamefont {Gefen}}, \
  and\ \bibinfo {author} {\bibfnamefont {A.}~\bibnamefont {Shnirman}},\ }\href
  {\doibase 10.1103/PhysRevB.99.045429} {\bibfield  {journal} {\bibinfo
  {journal} {Phys. Rev. B}\ }\textbf {\bibinfo {volume} {99}},\ \bibinfo
  {pages} {045429} (\bibinfo {year} {2019}{\natexlab{a}})}\BibitemShut
  {NoStop}%
\bibitem [{\citenamefont {Ludwig}\ \emph
  {et~al.}(2019{\natexlab{b}})\citenamefont {Ludwig}, \citenamefont
  {Burmistrov}, \citenamefont {Gefen},\ and\ \citenamefont
  {Shnirman}}]{ludwig2019-cng}%
  \BibitemOpen
  \bibfield  {author} {\bibinfo {author} {\bibfnamefont {T.}~\bibnamefont
  {Ludwig}}, \bibinfo {author} {\bibfnamefont {I.~S.}\ \bibnamefont
  {Burmistrov}}, \bibinfo {author} {\bibfnamefont {Y.}~\bibnamefont {Gefen}}, \
  and\ \bibinfo {author} {\bibfnamefont {A.}~\bibnamefont {Shnirman}},\
  }\href@noop {} {\enquote {\bibinfo {title} {Current noise geometrically
  generated by a driven magnet},}\ } (\bibinfo {year} {2019}{\natexlab{b}}),\
  \Eprint {http://arxiv.org/abs/1906.02730} {arXiv:1906.02730} \BibitemShut
  {NoStop}%
\bibitem [{\citenamefont {Tinkham}(2004)}]{tinkham2004introduction}%
  \BibitemOpen
  \bibfield  {author} {\bibinfo {author} {\bibfnamefont {M.}~\bibnamefont
  {Tinkham}},\ }\href@noop {} {\emph {\bibinfo {title} {Introduction to
  superconductivity}}}\ (\bibinfo  {publisher} {Courier Corporation},\ \bibinfo
  {year} {2004})\BibitemShut {NoStop}%
\bibitem [{\citenamefont {Hatami}\ \emph {et~al.}(2007)\citenamefont {Hatami},
  \citenamefont {Bauer}, \citenamefont {Zhang},\ and\ \citenamefont
  {Kelly}}]{hatami2007-tst}%
  \BibitemOpen
  \bibfield  {author} {\bibinfo {author} {\bibfnamefont {M.}~\bibnamefont
  {Hatami}}, \bibinfo {author} {\bibfnamefont {G.~E.~W.}\ \bibnamefont
  {Bauer}}, \bibinfo {author} {\bibfnamefont {Q.}~\bibnamefont {Zhang}}, \ and\
  \bibinfo {author} {\bibfnamefont {P.~J.}\ \bibnamefont {Kelly}},\ }\href
  {\doibase 10.1103/PhysRevLett.99.066603} {\bibfield  {journal} {\bibinfo
  {journal} {Phys. Rev. Lett.}\ }\textbf {\bibinfo {volume} {99}},\ \bibinfo
  {pages} {066603} (\bibinfo {year} {2007})}\BibitemShut {NoStop}%
\bibitem [{\citenamefont {Virtanen}\ and\ \citenamefont
  {Heikkil{\"{a}}}(2017)}]{virtanen2017-spt}%
  \BibitemOpen
  \bibfield  {author} {\bibinfo {author} {\bibfnamefont {P.}~\bibnamefont
  {Virtanen}}\ and\ \bibinfo {author} {\bibfnamefont {T.~T.}\ \bibnamefont
  {Heikkil{\"{a}}}},\ }\href {\doibase 10.1103/PhysRevLett.118.237701}
  {\bibfield  {journal} {\bibinfo  {journal} {Phys. Rev. Lett.}\ }\textbf
  {\bibinfo {volume} {118}},\ \bibinfo {pages} {237701} (\bibinfo {year}
  {2017})}\BibitemShut {NoStop}%
\bibitem [{\citenamefont {Utsumi}\ and\ \citenamefont
  {Taniguchi}(2015)}]{utsumi2015-fts}%
  \BibitemOpen
  \bibfield  {author} {\bibinfo {author} {\bibfnamefont {Y.}~\bibnamefont
  {Utsumi}}\ and\ \bibinfo {author} {\bibfnamefont {T.}~\bibnamefont
  {Taniguchi}},\ }\href {\doibase 10.1103/PhysRevLett.114.186601} {\bibfield
  {journal} {\bibinfo  {journal} {Phys. Rev. Lett.}\ }\textbf {\bibinfo
  {volume} {114}},\ \bibinfo {pages} {186601} (\bibinfo {year}
  {2015})}\BibitemShut {NoStop}%
\bibitem [{\citenamefont {Andrieux}\ and\ \citenamefont
  {Gaspard}(2004)}]{andrieux04}%
  \BibitemOpen
  \bibfield  {author} {\bibinfo {author} {\bibfnamefont {D.}~\bibnamefont
  {Andrieux}}\ and\ \bibinfo {author} {\bibfnamefont {P.}~\bibnamefont
  {Gaspard}},\ }\href {\doibase http://dx.doi.org/10.1063/1.1782391} {\bibfield
   {journal} {\bibinfo  {journal} {J. Chem. Phys.}\ }\textbf {\bibinfo {volume}
  {121}},\ \bibinfo {pages} {6167} (\bibinfo {year} {2004})}\BibitemShut
  {NoStop}%
\bibitem [{\citenamefont {Giazotto}\ \emph {et~al.}(2006)\citenamefont
  {Giazotto}, \citenamefont {Heikkil{\"a}}, \citenamefont {Luukanen},
  \citenamefont {Savin},\ and\ \citenamefont
  {Pekola}}]{giazotto2006opportunities}%
  \BibitemOpen
  \bibfield  {author} {\bibinfo {author} {\bibfnamefont {F.}~\bibnamefont
  {Giazotto}}, \bibinfo {author} {\bibfnamefont {T.~T.}\ \bibnamefont
  {Heikkil{\"a}}}, \bibinfo {author} {\bibfnamefont {A.}~\bibnamefont
  {Luukanen}}, \bibinfo {author} {\bibfnamefont {A.~M.}\ \bibnamefont {Savin}},
  \ and\ \bibinfo {author} {\bibfnamefont {J.~P.}\ \bibnamefont {Pekola}},\
  }\href {\doibase 10.1103/RevModPhys.78.217} {\bibfield  {journal} {\bibinfo
  {journal} {Rev. Mod. Phys.}\ }\textbf {\bibinfo {volume} {78}},\ \bibinfo
  {pages} {217} (\bibinfo {year} {2006})}\BibitemShut {NoStop}%
\bibitem [{\citenamefont {Dynes}\ \emph {et~al.}(1984)\citenamefont {Dynes},
  \citenamefont {Garno}, \citenamefont {Hertel},\ and\ \citenamefont
  {Orlando}}]{dynes1984tunneling}%
  \BibitemOpen
  \bibfield  {author} {\bibinfo {author} {\bibfnamefont {R.~C.}\ \bibnamefont
  {Dynes}}, \bibinfo {author} {\bibfnamefont {J.~P.}\ \bibnamefont {Garno}},
  \bibinfo {author} {\bibfnamefont {G.~B.}\ \bibnamefont {Hertel}}, \ and\
  \bibinfo {author} {\bibfnamefont {T.~P.}\ \bibnamefont {Orlando}},\ }\href
  {\doibase 10.1103/PhysRevLett.53.2437} {\bibfield  {journal} {\bibinfo
  {journal} {Phys. Rev. Lett.}\ }\textbf {\bibinfo {volume} {53}},\ \bibinfo
  {pages} {2437} (\bibinfo {year} {1984})}\BibitemShut {NoStop}%
\bibitem [{\citenamefont {Chudnovskiy}\ \emph {et~al.}(2008)\citenamefont
  {Chudnovskiy}, \citenamefont {Swiebodzinski},\ and\ \citenamefont
  {Kamenev}}]{chudnovskiy2008-sts}%
  \BibitemOpen
  \bibfield  {author} {\bibinfo {author} {\bibfnamefont {A.~L.}\ \bibnamefont
  {Chudnovskiy}}, \bibinfo {author} {\bibfnamefont {J.}~\bibnamefont
  {Swiebodzinski}}, \ and\ \bibinfo {author} {\bibfnamefont {A.}~\bibnamefont
  {Kamenev}},\ }\href {\doibase 10.1103/PhysRevLett.101.066601} {\bibfield
  {journal} {\bibinfo  {journal} {Phys. Rev. Lett.}\ }\textbf {\bibinfo
  {volume} {101}},\ \bibinfo {pages} {066601} (\bibinfo {year}
  {2008})}\BibitemShut {NoStop}%
\bibitem [{\citenamefont {Basko}\ and\ \citenamefont
  {Vavilov}(2009)}]{basko2009-sdm}%
  \BibitemOpen
  \bibfield  {author} {\bibinfo {author} {\bibfnamefont {D.~M.}\ \bibnamefont
  {Basko}}\ and\ \bibinfo {author} {\bibfnamefont {M.~G.}\ \bibnamefont
  {Vavilov}},\ }\href {\doibase 10.1103/PhysRevB.79.064418} {\bibfield
  {journal} {\bibinfo  {journal} {Phys. Rev. B}\ }\textbf {\bibinfo {volume}
  {79}},\ \bibinfo {pages} {064418} (\bibinfo {year} {2009})}\BibitemShut
  {NoStop}%
\bibitem [{\citenamefont {Kiselev}\ \emph {et~al.}(2003)\citenamefont
  {Kiselev}, \citenamefont {Sankey}, \citenamefont {Krivorotov}, \citenamefont
  {Emley}, \citenamefont {Schoelkopf}, \citenamefont {Buhrman},\ and\
  \citenamefont {Ralph}}]{kiselev2003-mon}%
  \BibitemOpen
  \bibfield  {author} {\bibinfo {author} {\bibfnamefont {S.~I.}\ \bibnamefont
  {Kiselev}}, \bibinfo {author} {\bibfnamefont {J.~C.}\ \bibnamefont {Sankey}},
  \bibinfo {author} {\bibfnamefont {I.~N.}\ \bibnamefont {Krivorotov}},
  \bibinfo {author} {\bibfnamefont {N.~C.}\ \bibnamefont {Emley}}, \bibinfo
  {author} {\bibfnamefont {R.~J.}\ \bibnamefont {Schoelkopf}}, \bibinfo
  {author} {\bibfnamefont {R.~A.}\ \bibnamefont {Buhrman}}, \ and\ \bibinfo
  {author} {\bibfnamefont {D.~C.}\ \bibnamefont {Ralph}},\ }\href {\doibase
  10.1038/nature01967} {\bibfield  {journal} {\bibinfo  {journal} {Nature}\
  }\textbf {\bibinfo {volume} {425}},\ \bibinfo {pages} {380} (\bibinfo {year}
  {2003})}\BibitemShut {NoStop}%
\bibitem [{\citenamefont {Rippard}\ \emph {et~al.}(2004)\citenamefont
  {Rippard}, \citenamefont {Pufall}, \citenamefont {Kaka}, \citenamefont
  {Russek},\ and\ \citenamefont {Silva}}]{rippard2004-dci}%
  \BibitemOpen
  \bibfield  {author} {\bibinfo {author} {\bibfnamefont {W.~H.}\ \bibnamefont
  {Rippard}}, \bibinfo {author} {\bibfnamefont {M.~R.}\ \bibnamefont {Pufall}},
  \bibinfo {author} {\bibfnamefont {S.}~\bibnamefont {Kaka}}, \bibinfo {author}
  {\bibfnamefont {S.~E.}\ \bibnamefont {Russek}}, \ and\ \bibinfo {author}
  {\bibfnamefont {T.~J.}\ \bibnamefont {Silva}},\ }\href {\doibase
  10.1103/PhysRevLett.92.027201} {\bibfield  {journal} {\bibinfo  {journal}
  {Phys. Rev. Lett.}\ }\textbf {\bibinfo {volume} {92}},\ \bibinfo {pages}
  {027201} (\bibinfo {year} {2004})}\BibitemShut {NoStop}%
\bibitem [{\citenamefont {Nagamine}\ \emph {et~al.}(2006)\citenamefont
  {Nagamine}, \citenamefont {Maehara}, \citenamefont {Tsunekawa}, \citenamefont
  {Djayaprawira}, \citenamefont {Watanabe}, \citenamefont {Yuasa},\ and\
  \citenamefont {Ando}}]{nagamine2006-ura}%
  \BibitemOpen
  \bibfield  {author} {\bibinfo {author} {\bibfnamefont {Y.}~\bibnamefont
  {Nagamine}}, \bibinfo {author} {\bibfnamefont {H.}~\bibnamefont {Maehara}},
  \bibinfo {author} {\bibfnamefont {K.}~\bibnamefont {Tsunekawa}}, \bibinfo
  {author} {\bibfnamefont {D.~D.}\ \bibnamefont {Djayaprawira}}, \bibinfo
  {author} {\bibfnamefont {N.}~\bibnamefont {Watanabe}}, \bibinfo {author}
  {\bibfnamefont {S.}~\bibnamefont {Yuasa}}, \ and\ \bibinfo {author}
  {\bibfnamefont {K.}~\bibnamefont {Ando}},\ }\href {\doibase
  10.1063/1.2352046} {\bibfield  {journal} {\bibinfo  {journal} {Appl. Phys.
  Lett.}\ }\textbf {\bibinfo {volume} {89}},\ \bibinfo {pages} {162507}
  (\bibinfo {year} {2006})}\BibitemShut {NoStop}%
\bibitem [{\citenamefont {Fransson}\ and\ \citenamefont
  {Zhu}(2008)}]{fransson2008-sdt}%
  \BibitemOpen
  \bibfield  {author} {\bibinfo {author} {\bibfnamefont {J.}~\bibnamefont
  {Fransson}}\ and\ \bibinfo {author} {\bibfnamefont {J.-X.}\ \bibnamefont
  {Zhu}},\ }\href {\doibase 10.1088/1367-2630/10/1/013017} {\bibfield
  {journal} {\bibinfo  {journal} {New J. Phys.}\ }\textbf {\bibinfo {volume}
  {10}},\ \bibinfo {pages} {013017} (\bibinfo {year} {2008})}\BibitemShut
  {NoStop}%
\bibitem [{\citenamefont {Zhu}\ \emph {et~al.}(2004)\citenamefont {Zhu},
  \citenamefont {Nussinov}, \citenamefont {Shnirman},\ and\ \citenamefont
  {Balatsky}}]{zhu2004-nsd}%
  \BibitemOpen
  \bibfield  {author} {\bibinfo {author} {\bibfnamefont {J.-X.}\ \bibnamefont
  {Zhu}}, \bibinfo {author} {\bibfnamefont {Z.}~\bibnamefont {Nussinov}},
  \bibinfo {author} {\bibfnamefont {A.}~\bibnamefont {Shnirman}}, \ and\
  \bibinfo {author} {\bibfnamefont {A.~V.}\ \bibnamefont {Balatsky}},\ }\href
  {\doibase 10.1103/PhysRevLett.92.107001} {\bibfield  {journal} {\bibinfo
  {journal} {Phys. Rev. Lett.}\ }\textbf {\bibinfo {volume} {92}},\ \bibinfo
  {pages} {107001} (\bibinfo {year} {2004})}\BibitemShut {NoStop}%
\bibitem [{\citenamefont {Kamenev}(2011)}]{kamenevbook}%
  \BibitemOpen
  \bibfield  {author} {\bibinfo {author} {\bibfnamefont {A.}~\bibnamefont
  {Kamenev}},\ }\href@noop {} {\emph {\bibinfo {title} {Field theory of
  non-equilibrium systems}}}\ (\bibinfo  {publisher} {Cambridge University
  Press},\ \bibinfo {year} {2011})\BibitemShut {NoStop}%
\bibitem [{\citenamefont {Eilenberger}(1968)}]{eilenberger1968-tog}%
  \BibitemOpen
  \bibfield  {author} {\bibinfo {author} {\bibfnamefont {G.}~\bibnamefont
  {Eilenberger}},\ }\href {\doibase 10.1007/BF01379803} {\bibfield  {journal}
  {\bibinfo  {journal} {Z. Phys}\ }\textbf {\bibinfo {volume} {214}},\ \bibinfo
  {pages} {195} (\bibinfo {year} {1968})}\BibitemShut {NoStop}%
\end{thebibliography}%

\appendix

\section{Tunneling currents}
\label{app:tunn}

Calculation of the tunneling currents from the model~\eqref{eq:hamiltonian} in the main text
can be done with standard Green function approaches. \cite{bergeret12}
Assuming a spin and momentum independent matrix element ($W_{jj'}=W$),
the $k$-spin component of the spin
current to S reads:
\begin{align}
  \label{eq:Isktun}
  I_s^k &= \frac{G_T}{32}
  \int_{-\infty}^\infty\dd{\epsilon}
  \tr \frac{\sigma_k}{2} [
    (R \check{g}_{F} R^\dagger)_+\check{g}_{S}
    -
    \check{g}_{S} (R \check{g}_{F} R^\dagger)_-]^K,
\end{align}
where the superscript $K$ refers to the Keldysh component and \(G_T=\pi\nu_F\nu_S|W|^2\) is the normal state tunneling conductance. The charge
and energy currents can be obtained by replacing
$\sigma_k/2\mapsto\hat{\tau}_3$ and  $\sigma_k/2\mapsto\epsilon$ in
Eq.~\eqref{eq:Isktun}, respectively. 
Here, $\sigma_j$ and $\hat{\tau}_j$ are Pauli
matrices in the spin and Nambu spaces, with the basis
$(\psi_\uparrow,\psi_\downarrow,-\psi_\downarrow^\dagger,\psi_\uparrow^\dagger)$,
and $X_+(\epsilon,t)=\int\dd{t'}e^{i\epsilon (t-t')}X(t,t')$,
$X_-(\epsilon,t)=\int\dd{t'}e^{i\epsilon (t'-t)}X(t',t)$.  Moreover,
$\check{g}_{F/S}(\epsilon)=\frac{2i}{\pi\nu_{F/S}}\hat{\tau}_3\sum_{j}\check{G}_{F/S}(\epsilon,\vec{p}_j)$
are state-summed Keldysh Green's functions,
normalized by the total density of states (DOS) $\nu_{F/S}$ at Fermi level,
of the ferromagnet and the spin-split superconductor.  The rotation
matrix
\begin{equation}
\begin{split}
R=e^{-i\phi\sigma_z/2}&e^{-i\theta\sigma_y/2}e^{i\phi\sigma_z/2}\\
\times\; &e^{-i\int^t\dd{t}\dot{\phi}(1-\cos\theta)\sigma_z/2}e^{-iV\hat{\tau}_3t}
\end{split}
\end{equation}
contains the Euler angles of the time-dependent magnetization direction vector
($\vec{m}\cdot\vec{\sigma}=R\sigma_zR^\dagger$), a Berry phase factor,
and voltage bias $V$. 
The Berry phase appears from the
Green function \cite{flebus2017-tmm,shnirman2015-gqn} of the conduction electrons in F
following adiabatically the changing magnetization.
For a metallic ferromagnet,
$\hat{g}_F^{R}-\hat{g}_F^{A}\simeq2\sum_\pm(\hat{\tau}_3\pm\sigma_z)\frac{\nu_{F,\pm}}{\nu_F}$
and $\hat{g}^{K}=[\hat{g}^{R}-\hat{g}^{A}](1-2f_0(\epsilon))$, where
$\nu_{F,\uparrow/\downarrow}:=\nu_{F,\pm}$ are the densities of
states of majority/minority spins at the Fermi level and
$f_0(\epsilon)=(1+e^{\epsilon/T})^{-1}$ is the Fermi distribution
function.

Evaluating Eq.~\eqref{eq:Isktun} for the different currents produces
Eqs.~(\ref{eq:avgic}--\ref{eq:avgtauz}) in the main text, with
$N_{S/F,\sigma=\pm}=\frac{1}{2}\tr[\frac{1+\hat{\tau}_3}{2}\frac{1+\sigma\sigma_z}{2}(\hat{g}_{S/F}^R-\hat{g}_{S/F}^A)]$.

Beyond linear response \eqref{eq:onsager}, we find the second-order
contributions to the current and torque:
\begin{align}
  \label{eq:current-2nd_order}
  \delta^{(2)}\overline{I_c}
  &=
  -\frac{\alpha^-_{2,0}}{2}\biggl[\sin^2(\theta) \Bigl(V\Omega - \frac{P \cos\theta}{4}\Omega^2\Bigr)
  \\\nonumber
  &
  + P\cos(\theta) V^2\biggr] -P\cos(\theta) \frac{A}{2} \left(\frac{\delta T}{T}\right)^2 - B\frac{\delta T}{T} V
  \,,
  \\
  \label{eq:torque-2nd_order}
  \frac{\delta^{(2)}\overline{\tau_z}}{\sin^2(\theta)}
  &=
  \frac{\alpha_{2,0}^-}{4}\Bigl[V^2 - P\cos(\theta) V\Omega + \frac{3+\cos(\theta)}{8} \Omega^2\Bigr]
  \\\nonumber
  &
  +\frac{A}{4} \biggl(\frac{\delta T}{T}\biggr)^2 + \frac{B}{4} \frac{\delta T}{T} \Omega
  \,,
\end{align}
where $\alpha_{i,j}^\mp=-(G_T/2)\int_{-\infty}^\infty\dd{\epsilon}\epsilon^j{}[N_{S,+}(\epsilon)\mp N_{S,-}(\epsilon)]f_0^{(i)}(\epsilon)$,
and $A=2\alpha_{1,1}^- {+} \alpha_{2,2}^-$, $B=\alpha_{1,0}^+{+}\alpha_{2,1}^+$.

For \(\Omega\ll \Delta\), the onset of the voltage driven spin oscillations [Fig.~\ref{fig:spintorque}(a)] can be determined from Eqs.~\eqref{eq:onsager} and \eqref{eq:torque-2nd_order} to occur at $V_o=\pm 4\sqrt{e^2 \mathcal S A_{\rm eff} \Omega/\alpha^-_{2,0}}$.

In addition to the spin transfer torque (STT) discussed in the main
text, the electron transfer between F and the spin-split S generates
also other torque components acting on $F$. This effect can be found
from Eq.~\eqref{eq:Isktun}, and appears in the torque components
$\overline{\tau_{x/y}}$ perpendicular to the equilibrium magnetization
$\hat{z}$.

In the main text, we neglect these torques, because any equilibrium
torques can be absorbed to a renormalization of the effective magnetic
field, and moreover, in the limit of weak damping and torques the
components perpendicular to $\hat{z}$ such that
$\tau_{x/y}\ll\mathcal{S}\Omega$ have little effect on the
dynamics. In contrast, the component in the main text has a significant effect
already at $\overline{\tau_z}\sim{}A_0\mathcal{S}\Omega\ll\mathcal{S}\Omega$.

For completeness, we write here the expressions for all torques, as
obtained from Eq.~\eqref{eq:Isktun}.  Equation~\eqref{eq:avgtauz} in
the main text gives  the dissipative contribution to $\tau_z$. Similar
contributions can be found for $\tau_{x/y}$:
\begin{align}
  \label{eq:tau-xy}
  \overline{\tau_{x/y}}
  &=
  -
  \frac{G_T}{8}
  \int_{-\infty}^\infty\dd{\epsilon}\sum_{\sigma\sigma'}  \frac{(1+\sigma\sigma'\cos\theta)^2}{2}
  N_{S,x/y}
  \\\notag&\qquad
  \times[f_F(\epsilon-\Omega_{\sigma\sigma'}-V)-f_S(\epsilon)]
  \,,
\end{align}
where $N_{S,0/x/y/z}=\frac{1}{2}\tr\frac{1+\tau_3}{2}\frac{\sigma_{0/x/y/z}}{2}(\hat{g}_{S}^R-\hat{g}_S^A)$.

In addition, there are two remaining contributions, the equilibrium spin
torque, and a Kramers--Kronig counterpart to the density of states
term. Terms of the latter type commonly appear in calculations of
time-dependent response.  To find it, we need
$\hat{g}^{R+A}=\hat{g}^{R}+\hat{g}^{A}$.  We can evaluate them e.g. in
a model with a parabolic spectrum in 3D, $\xi_k=k^2/(2m) - \mu$.  In
the superconductor, $h,\Delta \ll\mu_S$ and in the magnet,
$\Delta=0$. Evaluating the momentum sum yields
\begin{align}
  \hat{g}_S^{R+A}
  &\overset{\mu_S\to\infty}{\simeq}
  \hat{g}^R_{S,\rm qcl} + \hat{g}^A_{S,\rm qcl}
  +
  \hat{g}_{F}^{R+A}\rvert_{\vec{h}_F\mapsto\vec{h},\mu_F\mapsto\mu_S}
  \,,
  \\\notag
  \hat{g}_F^{R+A}
  &=
  2i
  a
  \Re
  \sqrt{-[(\epsilon - h_F\sigma_z)\hat{\tau}_3 + \mu_F]/|\mu_F|}
  +
  C
  \,.
\end{align}
Here $\hat{g}^{R/A}_{S,\rm qcl}$, are quasiclassical low-energy Green
functions \cite{eilenberger1968-tog},
$1/a=\sum_\pm\sqrt{1\pm{}h_F/\mu_F}$, and
$h_F=\frac{\nu_{F\downarrow}^2-\nu_{F\uparrow}^2}{\nu_{F\downarrow}^2+\nu_{F\uparrow}^2}\mu_F$
the internal exchange field in F in the model. Moreover, $C$ are
scalars independent of $\epsilon$, $\vec{h}$, and $\Delta$, and drop
out from expressions for the observables here.  

Neglecting terms of
order $\Delta/\mu,T/\mu,\Omega/\mu$, we find the remaining terms in the
spin current,
\begin{align}
  \label{eq:fm-nonfermi-torque}
  \vec{I}_S''
  &=
  \vec{I}_{S,\rm eq}'' + \delta\vec{I}_{S}''
  \\
  \delta\vec{I}_{S}''
  &=
  -
  \frac{G_T}{64}
  \int_{-\infty}^\infty\dd{\epsilon}
  \sum_{\sigma\sigma'}
  \tanh\frac{\epsilon - \Omega_{\sigma\sigma'} - V}{2T_F}
  \\\notag&\qquad\times
  (\sigma\hat{z}+\sigma'\vec{m}(t))\times\vec{P}(\epsilon)N_{F,\sigma'}
  \,,
\end{align}
where
$\vec{P}(\epsilon)=\frac{1}{2i}\tr\frac{1+\tau_3}{2}\vec{\sigma}[\hat{g}^R_{S,\rm{}qcl}(\epsilon) + \hat{g}^A_{S,\rm{}qcl}(\epsilon)]$. It has the
symmetry $\vec{P}(-\epsilon)=\vec{P}(\epsilon)$.
For a BCS superconductor, the integrand is nonzero only inside the gap,
$|\epsilon\pm{}h|<\Delta$.

The equilibrium spin current $\vec{I}_{S,\rm eq}''$ is
related to the exchange coupling between F and FI mediated by the
electrons in the superconductor.  It can be absorbed to a
small renormalization of the effective magnetic field acting on F.  While
its value can be calculated
in the above tunneling model, the model is
not sufficient for describing this non-Fermi surface term in the
realistic situation. The superconducting correction
$\delta\vec{I}_{S}''$ vanishes at equilibrium, but may contribute to
nonequilibrium response. This torque however has $\overline{\tau_z''}=0$
and can be neglected similarly as in Eq.~\eqref{eq:tau-xy}.

\section{Adiabatic Green function}
\label{app:green}

In the tunneling calculation of Eq.~\ref{eq:Isktun}, an expression
for the adiabatic Green function of the electrons on the ferromagnet
with dynamic magnetization appears. For completeness, we discuss its meaning here.
The nonequilibrium Green function for free electrons
in a time-dependent exchange field,
$H(t)=\sum_{n\sigma\sigma'}c^\dagger_{n\sigma}[\mathcal{H}_n(t)]_{\sigma\sigma'}c_{n\sigma'}$,
$\mathcal{H}_n(t)=\epsilon_n+\vec{h}(t)\cdot\vec{\sigma}$, with
a thermal initial state at $t=0$ is
$G_{n}^>(t,t')=-iU_n(t,0)(1 - \rho_n)U_n(0,t')^\dagger$,
where $i\partial_tU_n(t,t') = [\epsilon_n -
  \vec{h}(t)\cdot\vec{\sigma}]U_n(t,t')$, $U(t,t)=1$, and $\rho_n=[1 +
  e^{\mathcal{H}_n(0)/T}]^{-1}$. In an adiabatic approximation for
$|\dot{h}|\ll{}h^2$,
$U_n(t,t')\simeq{}e^{-i(t-t')\epsilon_n}R(t)e^{i\varphi_n(t,t')\sigma_z/2}R(t')^\dagger$,
where $R(t)\sigma_zR(t)^\dagger=\vec{h}(t)\cdot\vec{\sigma}$ and
$\varphi_n(t,t')=i\int_{t'}^t\dd{t''}\tr \sigma_z
R(t'')^\dagger\partial_{t''}R(t'')$.  In terms of Euler angles
$\vec{h}=(\cos\phi\sin\theta,\sin\phi\sin\theta,\cos\theta)$ we write
$R=e^{-i\phi\sigma_z/2}e^{-i\theta\sigma_y/2}e^{i\phi\sigma_z/2}e^{-i\chi\sigma_z/2}$.
The function $\chi(t)$ is arbitrary, but $U_n$ does not depend on it.
For simplicity, we choose $\chi=\int^t\dd{t'}\dot{\phi}(1-\cos\theta)$, which gives
$\varphi_n=0$. With this choice, the adiabatic Green function becomes
\begin{align}
  G^>_n(t,t')=R(t)G_{n,0}^>(t-t')R(t')^\dagger
  \,,
\end{align}
and the electron Berry phase appears only in the rotation matrix.
This is equivalent to the ``rotating frame'' picture
used in the main text and other works \cite{tserkovnyak2005-nmd,tserkovnyak2008-tbe}.

\end{document}